\documentstyle[twocolumn,twoside]{article}
\newcommand{\FF}{{\cal F \hspace*{-2ex} F}}
\newcommand{\jj}{\widetilde{\mbox{\bf \j}}}
\newcommand{\Lie}{\pounds \hspace*{-1.7ex} \pounds}
\title{\Large \bf Five-Dimensional Tangent Vectors in Space-Time \\
\large \bf II. Differential-Geometric Approach}
\author{Alexander Krasulin \\ \it Institute for Nuclear Research of the
Russian Academy of Sciences \\ \it 60th October Anniversary Prospect, 7a,
117312 Moscow, Russia}
\date{\normalsize \bf Abstract \\ \mbox{ } \\ \begin{minipage}{400pt}
\normalsize In this part of the series five-dimensional tangent vectors
are introduced first as equivalence classes of parametrized curves and
then as differential-algebraic operators that act on scalar functions.
I then examine their basic algebraic properties and their parallel transport
in the particular case where space-time possesses a special local symmetry.
After that I give definition to five-dimensional tangent vectors associated
with dimensional curve parameters and show that they can be identified with
the five-vectors introduced formally in part I. In conclusion I speak about
differential forms associated with five-vectors. \end{minipage} }
\begin{document}

\maketitle

\begin{flushleft} \bf
1. Five-vectors as equivalence classes \\
\hspace{2.5ex} of parametrized curves \\
\vspace{2ex} \rm A. \it Definition \\
\end{flushleft}
Consider a set $\Re$ of all smooth parametrized curves going through a fixed
space-time point $Q$. I will lable these curves with calligraphic capital
Roman letters: $\cal A, \, B, \, C,$ etc. The parameter of curve $\cal A$
will be denoted as $\lambda_{\cal A}$.

If $f$ is a real scalar function defined in the vicinity of $Q$,
one can evaluate its derivative at $Q$ along a given curve $\cal A$:
\begin{displaymath}
\left. \frac{d f(P(\lambda_{\cal A}))}{d \lambda_{\cal A}}
\right| _{\lambda_{\cal A} = \lambda_{\cal A}(Q)},
\end{displaymath}
and I will denote this derivative as $\partial_{\cal A} f |_Q$.

Let us focus our attention on the behaviour of curves in the infinitesimal
vicinity of $Q$. From that point of view, $\Re$ can be divided into classes
of equivalent curves that coincide in direction or in direction and
parametrization. One can consider three degrees to which two given curves,
$\cal A$ and $\cal B$, may coincide:
\begin{enumerate}
   \item The two curves come out of $Q$ in the same direction. A more precise
   formulation is the following: there exists a real positive number $a$ such
   that for any scalar function $f$
\begin{equation}
\partial_{\cal A} f |_{Q} = a \cdot \partial_{\cal B} f |_{Q}.
\end{equation}
   \item The two curves come out of $Q$ in the same direction and in the
   vicinity of $Q$ their parameters change with equal rates. More precisely:
   for any scalar function $f$
\begin{equation}
\partial_{\cal A} f |_{Q} = \partial_{\cal B} f |_{Q}.
\end{equation}
   \item The two curves come out of $Q$ in the same direction; their
   parameters change with equal rates in the vicinity of $Q$; and the
   values of these parameters at $Q$ are the same. This means that
\begin{flushright}
\hspace*{4ex} \hfill $\lambda_{\cal A}(Q) = \lambda_{\cal B}(Q)$
\hfill {\rm (3a)}
\end{flushright}
and for any scalar function $f$
\begin{flushright}
\hspace*{4ex} \hfill $\partial_{\cal A} f |_{Q} = \partial_{\cal B} f |_{Q}.$
\hfill {\rm (3b)}
\end{flushright} \setcounter{equation}{3} \end{enumerate}

It is a simple matter to check that relations (1), (2) and (3) are all
equivalence relations on $\Re$, and for each of them one can consider the
corresponding quotient set---the set whose elements are classes of equivalent
curves.

Relation (1) is of no interest to us and I will not consider it any further.

The elements of the quotient set corresponding to relation (2) will be
denoted with capital boldface Roman letters: $\bf A$, $\bf B$, $\bf C$, etc.
According to relation (2), the derivative of any scalar function $f$ at $Q$
is the same for all curves belonging to a given class $\bf A$, so it makes
sense to introduce the notation $\partial_{\bf A} f|_{Q}$.

In a natural way, one can define the addition of two equivalence classes
$\bf A$ and $\bf B$: $\bf A + B$ is such an equivalence class that for any
scalar function $f$ one has
\begin{displaymath}
\partial_{\bf A+B} f|_{Q} = \partial_{\bf A} f|_{Q} + \partial_{\bf B} f|_{Q}.
\end{displaymath}
It is easy to prove that such a sum exists for any pair of equivalence
classes.

In a similar manner one can give definition to the product of an equivalence
class $\bf A$ and a real number $k$: $k{\bf A}$ is such an equivalnce class
that
\begin{displaymath}
\partial_{k{\bf A}} f|_{Q} = k \cdot \partial_{\bf A} f|_{Q}
\end{displaymath}
for any scalar function $f$. Again, one can verify that $k{\bf A}$ exists
for any $\bf A$ and any $k$.

With thus defined addition and multiplication by a real number, the set of
all equivalence classes corresponding to relation (2) becomes a vector space.
This space is four-dimensional, and I will denote it as $V_{4}$. As it will
be discussed in section 2, the elements of $V_{4}$ can be identified with
four-dimensional tangent vectors, so in the following I will refer to them
as to {\em four-vectors}.

Let us now turn to the quotient set associated with relation (3). Its
elements will be denoted with lower-case boldface Roman letters: $\bf a$,
$\bf b$, $\bf c$, etc. As in the case of four-vectors, one can introduce the
notation $\partial_{\bf a} f|_{Q}$ for the common value of the derivatives of
any scalar function $f$ along all the curves belonging to a given equivalence
class $\bf a$. Similarly, the common value of the parameters of all these
curves at $Q$ will be denoted as $\lambda_{\bf a} (Q)$.

One can now give the following definitions to the sum of two equivalence
classes $\bf a$ and $\bf b$ and to the product of an equivalence class
$\bf a$ and a real number $k$: $\bf a+b$ and $k{\bf a}$ are such equivalence
classes that
\begin{displaymath} \begin{array}{l}
\lambda_{\bf a+b} (Q) = \lambda_{\bf a} (Q) + \lambda_{\bf b} (Q), \\
\lambda_{k{\bf a}} (Q) = k \cdot \lambda_{\bf a} (Q)
\end{array} \end{displaymath}
and for any scalar function $f$
\begin{displaymath} \begin{array}{l}
\partial_{\bf a+b}f|_{Q}=\partial_{\bf a}f|_{Q}+\partial_{\bf b}f|_{Q}, \\
\partial_{k{\bf a}} f|_{Q} = k \cdot \partial_{\bf a} f|_{Q}.
\end{array} \end{displaymath}
One can easily check that such a sum and such a product exist respectively
for any two equivalence classes and for any equivalence class and any real
number. These two operations turn the quotient set associated with relation
(3) into a vector space, whose dimension is evidently five. Let us denote
this space as $V_{5}$ and call its elements {\em five-dimensional tangent
vectors} or simply {\em five-vectors}. In section 2 I will consider another,
equivalent representation for these vectors and later on will show that
they have all the formal properties of those five-vectors that have been
introduced in part I.

\begin{flushleft}
B. \it Structure of the five-vector space
\end{flushleft}
As any other vector space, $V_{5}$ is completely isotropic with respect to
its two composition laws and has no distinguished direction nor any other
distinguished subspace of nonzero dimension. However, one {\em can}
distinguish two subspaces in $V_{5}$ by associating them with certain
classes of parametrized curves.

Let us consider all those curves from $\Re$ for which $\partial f|_{Q} = 0$
for any scalar function $f$. It is evident that all these curves belong to
the same equivalence class with respect to relation (2) and that this class
is the zero vector in $V_{4}$. With respect to relation (3), the considered
curves belong to equivalence classes that make up a one-dimensional subspace
in $V_{5}$, which I will denote as $\cal E$. One can say that $\cal E$ is
made up by all those five-vectors that do not correspond to any direction
in the manifold.

Another distinguished subspace in $V_{5}$ can be obtained by considering
all those curves from $\Re$ for which $\lambda(Q) = 0$. The four-vectors
corresponding to these curves are all the vectors of $V_{4}$. The
corresponding five-vectors make up a four-dimensional subspace in $V_{5}$,
which I will denote as $\cal Z$. It is easy to see that $\cal E$ and
$\cal Z$ have only one common element---the zero vector, and that $V_{5}$
is the direct sum of $\cal E$ and $\cal Z$. The components of an
arbitrary five-vector $\bf u$ in these two subspaces will be denoted as
${\bf u}^{\cal E}$ and ${\bf u}^{\cal Z}$, respectively.

Other properties of $\cal E$ and $\cal Z$ will be discussed below.

\vspace{2ex} \begin{flushleft}
C. \it Relation between four- and five-vectors
\end{flushleft}
As it follows from the definition of four- and five-vectors given above,
there exists a set-theoretic relation between $V_{4}$ and $V_{5}$: the former
is the quotient set corresponding to the following equivalence relation on
$V_{5}$:
\begin{displaymath}
{\bf a} \equiv {\bf b} \Leftrightarrow \partial_{\bf a} f|_{Q} =
\partial_{\bf b} f|_{Q} \mbox{ for any scalar function } f.
\end{displaymath}
Denoting this relation as $R$, one has $V_{4} = V_{5}/R$. The fact that
$\bf A$ is the equivalence class of $\bf a$ will be denoted as ${\bf a}
\in {\bf A}$. From the definition of symbols $\partial_{\bf a}$ and
$\partial_{\bf A}$ it follows that ${\bf a} \in {\bf A}$ if and only if
$\partial_{\bf a} = \partial_{\bf A}$. It is a simple matter to see that
$R$ has the following linearity properties: if ${\bf a} \equiv {\bf b} \,
(\bmod R)$ and ${\bf c} \equiv {\bf d} \, (\bmod R)$, then ${\bf a+c} \equiv
{\bf b+d} \, (\bmod R)$ and $k{\bf a} \equiv k{\bf b} \, (\bmod R)$, where
$k$ is an arbitrary real number. Thus, as any other equivalence relation
with such properties, $R$ can be presented in the following form:
\begin{displaymath}
{\bf a} \equiv {\bf b} \, (\bmod R) \Leftrightarrow  {\bf a}-{\bf b} \in W,
\end{displaymath}
where $W$ is the subspace in $V_{5}$ that contains all the five-vectors
equivalent to the zero vector. It is easy to see that $W$ coincides with the
one-dimensional subspace $\cal E$ introduced in the previous subsection, so
$R$ can be reformulated as:
\begin{displaymath}
{\bf a} \equiv {\bf b} \, (\bmod R) \Leftrightarrow
{\bf a} = {\bf b} + {\bf e}, \mbox{ where } {\bf e} \in {\cal E}.
\end{displaymath}
The latter condition is equivalent to $\bf a$ and $\bf b$ having equal
components in the four-dimensional subspace $\cal Z$ or, for that matter,
in any subspace complementary to $\cal E$. This means that there exists a
one-to-one correspondence between the five-vectors from $\cal Z$ and
four-vectors, and this correspondence is evidently a homomorphism.

Let me say a few words about the selection of bases in $V_{4}$ and $V_{5}$
and their transformation.

A typical five-vector basis will be denoted as ${\bf e}_{A}$, where $A$
(as all capital latin indices) runs 0, 1, 2, 3, and 5. One can choose a
basis in $V_{5}$ arbitrarily, but it is more convenient to select the fifth
basis vector belonging to $\cal E$. Such bases will be called {\em standard}
and will be used in all calculations.

The basis in $V_{4}$ can be chosen arbitrarily and independently of the basis
in $V_{5}$. It is more convenient though to associate it with the five-vector
basis. A natural choice is to take ${\bf E}_{\alpha}$ to be the equivalence
classes of the basis five-vectors ${\bf e}_{\alpha}$ (the equivalence class
of ${\bf e}_{5}$ is the zero four-vector). I will refer to this basis as to
the one {\em associated} with the basis ${\bf e}_{A}$ in $V_{5}$.

If ${\bf e}_{A}$ and ${\bf e}'_{A}$ are two standard bases in $V_{5}$ and
${\bf e}'_{A} = {\bf e}_{B} L^{B}_{\, A}$, then $L^{B}_{\, A}$ can be shown
to satisfy the condition
\begin{displaymath}
L^{\alpha}_{\, 5} = 0 \; \, {\rm for \; all} \; \, \alpha.
\end{displaymath}
The corresponding equivalence classes are related as ${\bf E}'_{\alpha} =
{\bf E}_{\beta} L^{\beta}_{\; \alpha}$.

\vspace{2ex} \begin{flushleft}
D. \it Reminder on the inner product of four-vectors
\end{flushleft}
Four-vectors inherit their inner product from the Riemannian metric of
space-time. The latter is a rule that assigns a certain number, called
interval, to each finite continuous line. This number is additive, and for
an infinitesimal line connecting two points with coordinates $x^{\alpha}$ and
$x^{\alpha} + dx^{\alpha}$ it equals
\begin{equation}
\sqrt{g_{\alpha \beta}(x) dx^{\alpha} dx^{\beta}} \, + \mbox{ terms of
higher order in } dx,
\end{equation}
where $g_{\alpha \beta}$ is a real nondegenerate $4 \times 4$ matrix with the
signature $(+,-,-,-)$.

Consider now a parametrized curve coming out of a point $Q$. According to
formula (4), the interval assigned to the part of the curve between $Q$ and
a nearby point corresponding to the parameter value $\lambda(Q) + d \lambda$
is
\begin{displaymath} \left. \begin{array}{l}
\sqrt{g_{\alpha \beta}(Q) (\partial x^{\alpha}/\partial \lambda)_{Q}
(\partial x^{\beta}/\partial \lambda)_{Q}} \cdot d \lambda \\
\hspace{10ex} \rule{0ex}{3ex} + \mbox{ terms of higher order in } d \lambda.
\end{array} \right. \end{displaymath}
Since $(\partial x^{\alpha}/\partial \lambda)_{Q}$ is the same for all curves
from a given equivalence class associated with relation (2), the expression
under the radical sign is a function of the four-vector corresponding to the
curve rather than of the curve itself. This enables one to assign a number to
each four-vector, which is interpreted as its length squared. More precisely,
the inner product $g$ is defined as a real bilinear symmetric function of
two four-vectors such that for any four-vector $\bf U$
\begin{displaymath}
g({\bf U,U}) = g_{\alpha \beta}(Q) (\partial_{\bf U} x^{\alpha})_{Q}
(\partial_{\bf U} x^{\beta})_{Q}.
\end{displaymath}

The interval is a dimensional quantity. It is measured in centimeters or
seconds or in any other units of length or time. Accordingly, the quantity
under the radical sign in formula (4) is measured in $cm^{2}$ or $sec^{2}$
or in some other squared units. Throughout sections 1 and 2 of this paper
I will consider only dimensionless coordinates and curve parameters. Then,
if the interval is measured, say, in centimeters, the elements of the matrix
$g_{\alpha \beta}$ will be measured in $cm^{2}$, $g^{\alpha \beta}$ will be
measured in $cm^{-2}$, and the connection coefficients for four- and
five-vector fields will be dimensionless.

\vspace{2ex} \begin{flushleft}
E. \it Symmetries
\end{flushleft}
The set $\Re$ of all parametrized curves going through an arbitrary point
$Q$ has a certain symmetry with respect to the behaviour of curves in the
infinitesimal vicinity of $Q$. Namely, there exist certain maps of
$\Re$ onto itself that have the following properties:
\begin{enumerate}
  \item If ${\cal A} \mapsto {\cal A}'$, then $\lambda_{\cal A}(Q) =
   \lambda_{\cal A'}(Q)$.
  \item If ${\cal A} \mapsto {\cal A}'$ and ${\cal B} \mapsto {\cal B}'$, and
   for any scalar function $f$ one has $\partial_{\cal A} f|_{Q} = k \cdot
   \partial_{\cal B} f|_{Q}$, where $k$ is some constant factor, then for any
   scalar function $f$ one has $\partial_{\cal A'} f|_{Q} = k \cdot
   \partial_{\cal B'} f|_{Q}$.
  \item If ${\cal A} \mapsto {\cal A}'$, ${\cal B} \mapsto {\cal B}'$, and
   ${\cal C} \mapsto {\cal C}'$, and for any scalar function $f$ one has
   $\partial_{\cal A}f|_{Q}+\partial_{\cal B}f|_{Q}=\partial_{\cal C}f|_{Q}$,
   then for any scalar function $f$ one has $\partial_{\cal A'} f|_{Q} +
   \partial_{\cal B'} f|_{Q} = \partial_{\cal C'} f|_{Q}$.
  \item If ${\cal A} \mapsto {\cal A}'$, then
\begin{displaymath} \left. \begin{array}{l}
g_{\alpha \beta}(Q) (\partial_{\cal A} x^{\alpha})_{Q} (\partial_{\cal A}
x^{\beta})_{Q} \\ \hspace{15ex} \rule{0ex}{3ex} = g_{\alpha \beta}(Q)
(\partial_{\cal A'} x^{\alpha})_{Q} (\partial_{\cal A'} x^{\beta})_{Q}.
\end{array} \right. \end{displaymath} \end{enumerate}

Property 2 at $k=1$ means that such transformations of $\Re$ induce maps
of $V_{4}$ onto itself. Properties 2 and 3 mean that these transformations
of $V_{4}$ are linear, and property 4 means that they conserve the inner
product of two four-vectors.

Property 1 and property 2 at $k=1$ mean that the considered transformations of
$\Re$ also induce maps of $V_{5}$ onto itself. Properties 1, 2, and 3
mean that these maps are linear. Property 1 means that a vector from
$\cal Z$ is transformed into a vector from $\cal Z$. And properties 1 and 2
mean that vectors from $\cal E$ are not changed at all.

Let us now find the corresponding transformation matrices for $V_{4}$ and
$V_{5}$.

Let ${\bf E}_{\alpha}$ be an arbitrary orthonormal basis in $V_{4}$ and let
us take that under the considered transformation these basis vectors are
transformed into ${\bf E}'_{\alpha} = {\bf E}_{\beta}
\Lambda^{\beta}_{\; \alpha}$. Since the transformation should conserve
the inner product, and the basis ${\bf E}_{\alpha}$ is orthonormal,
$\Lambda^{\beta}_{\; \alpha}$ should be a matrix from O(3,1). As a basis in
$V_{5}$ let us take a standard basis where ${\bf e}_{\alpha} \in {\cal Z}$
and ${\bf e}_{\alpha} \in {\bf E}_{\alpha}$. Let us suppose that
${\bf e}_{A}$ are transformed into ${\bf e}'_{A} = {\bf e}_{B}L^{B}_{\; A}$,
where $L^{B}_{\; A}$ is some real nondegenerate $5 \times 5$ matrix. Since
vectors from $\cal E$ do not change under the considered transformation, one
should have $L^{5}_{\, 5} = 1$ and $L^{\alpha}_{\, 5} = 0$ for all $\alpha$.
Since vectors from $\cal Z$ are transformed into vectors from $\cal Z$, one
should have $L^{5}_{\, \alpha} = 0$ for all $\alpha$. Finally, owing to the
one-to-one correspondence between $\cal Z$ and $V_{4}$, one should have
$L^{\alpha}_{\, \beta} = \Lambda^{\alpha}_{\, \beta} \in$ O(3,1).

\vspace{2ex} \begin{flushleft}
F. \it Inner product of five-vectors
\end{flushleft}
The method used in subsection D to define the inner product $g$ for
four-vectors is also applicable in the case of five-vectors. The resulting
inner product on $V_{5}$, which for the time being I will denote as $h'$,
is a real bilinear symmetric function of two five-vectors such that for any
five-vector $\bf u$
\begin{displaymath}
h'({\bf u,u}) = g_{\alpha \beta}(Q) (\partial_{\bf u} x^{\alpha})_{Q}
(\partial_{\bf u} x^{\beta})_{Q}
\end{displaymath}
($Q$ is the space-time point where one considers the tangent space of
five-vectors). Since the value of the derivative $\partial_{\bf u}$ is the
same for all five-vectors corresponding to the same four-vector, $h'$ will
be a degenerate inner product. It is not difficult to see that the subspace
of all degenerate five-vectors for $h'$ (of all such five-vectors $\bf u$
that $h'({\bf u,v}) = 0$ for any $\bf v$) coincides with $\cal E$ and that
$h'$ is nondegenerate within any subspace complementary to $\cal E$. It is
also apparent that for any $\bf u$ and $\bf v$ one has
\begin{equation}
h'({\bf u,v}) = g({\bf U,V}),
\end{equation}
where $\bf u \in U$ and $\bf v \in V$.

It is not difficult to construct from $h'$ a nondegenerate inner product on
$V_{5}$. For that one should consider another natural measure that exists for
five-vectors: to each five-vector $\bf u$ one can put into correspondence
the value of the relevant curve parameter, $\lambda_{\bf u}$. If one then
interprets this latter number as the length of vector $\bf u$, one will
obtain another inner product---let us denote it as $h''$---which will also
be degenerate. It is easy to see that $h''({\bf u,v}) = \lambda_{\bf u} \cdot
\lambda_{\bf v}$. Consequently, the subspace of all degenerate vectors for
$h''$ coincides with $\cal Z$ and $h''$ is nondegenerate within any
(one-dimensional) subspace complementary to $\cal Z$.

One should now notice that the subspaces of degenerate vectors for $h'$ and
$h''$ are complementary to each other, which means that the sum of $h'$ and
$h''$ will be a nondegenerate inner product on $V_{5}$. The only problem in
constructing such a sum is that $h'$ is a dimensional quantity and is measured
in the same units as $g$, whereas $h''$, being the product of curve
parameters, does not have a dimension. We thus see that to construct a
nondegenerate inner product on $V_{5}$ from $h'$ and $h''$, one needs a
dimensional constant, $\xi$, which would play a role similar to that of the
speed of light: it would establish a relation between different units used to
measure the same quantity. The resulting inner product measured in the same
units as $g$ will be
\begin{equation}
h({\bf u,v}) = h'({\bf u,v}) + \xi \cdot h''({\bf u,v}).
\end{equation}

The same result can be obtained from considerations of another kind. For
that one should adopt the view-point that four-vectors and five-vectors
are subordinate objects, whose algebraic properties are determined by the
properties of the manifold with which they are associated. In particular,
this means that the structure of $V_{5}$ should have a symmetry no less than
the symmetry of $\Re$. This, in its turn, means that {\em any} inner product
of five-vectors should be invariant under the transformations discussed in
the previous subsection.

Let us consider the same five-vector basis ${\bf e}_{A}$ that has been used
in subsection E. It is a simple matter to show that the matrix $h_{AB}
\equiv h({\bf e}_{A},{\bf e}_{B})$ of any nondegenerate inner product $h$
satisfying the above symmetry requirement has to be of the form
\begin{equation}
h_{\alpha \beta} = a \cdot \eta_{\alpha \beta}, \; h_{\alpha 5}
= h_{5 \alpha} = 0, \; h_{55} = b,
\end{equation}
where $a$ and $b$ are some nonzero constants. A direct consequence of
these formulae is that any five-vector from $\cal Z$ is orthogonal to
any five-vector from $\cal E$, so for any $\bf u$ and $\bf v$
\begin{equation}
h({\bf u,v}) = h({\bf u^{\cal Z},v^{\cal Z}})+h({\bf u^{\cal E},v^{\cal E}}).
\end{equation}
Another consequence of formulae (7) is that the inner product of any two
five-vectors from $\cal Z$ is proportional to the inner product of the
corresponding four-vectors. Thus, if the overall normalization of $h$ is
selected in such a way that the proportionality factor between $h$ and $g$
be unity, one will have
\begin{displaymath}
h({\bf u^{\cal Z},v^{\cal Z}}) = g({\bf U,V}) = h'({\bf u,v}).
\end{displaymath}
Finally, one should observe that the $\cal E$-component of any five-vector
$\bf u$ equals $\lambda_{\bf u} \cdot {\bf i}$, where $\bf i$ is the vector
from $\cal E$ that corresponds to the unity value of the parameter:
$\lambda_{\bf i} = 1$. Consequently,
\begin{displaymath}
h({\bf u}^{\cal E},{\bf v}^{\cal E})  = \lambda_{\bf u} \lambda_{\bf v} \,
h({\bf i,i}) = h({\bf i,i}) \, h''({\bf u,v}),
\end{displaymath}
and formula (8) acquires the form of formula (6) with $\xi = h({\bf i,i})$.
Thus, at an appropriate choice of its overall normalization factor, any
nondegenerate inner product on $V_{5}$ satisfying the above, quite natural
symmetry requirement has the form indicated in formula (6).

It is obvious that constant $\xi$ is not determined by the Riemannian metric
of space-time nor by symmetry considerations, and consequently the same is
true of the nondegenerate inner product of five-vectors. This is a
distinctive feature of five-dimensional tangent vectors (and of similar
objects in other manifolds) and is a consequence of that specific way in
which five-vectors are associated with space-time.

In the previous subsection I have introduced a five-vector basis where
${\bf e}_{\alpha} \in {\cal Z}$. As we have seen above, in terms of the
five-vector inner product this means that all ${\bf e}_{\alpha}$ are
orthogonal to ${\bf e}_{5}$. This is one of the two conditions satisfied by
a {\em regular} five-vector basis defined in section 3 of part I within
the formal theory, the other condition being that $h({\bf e}_{5},{\bf e}_{5})
= 1$. When five-vectors are introduced as equivalence classes of parametrized
curves, it is more convenient to define the regular basis in a slightly
different way, equating to unity not the value of $h({\bf e}_{5},
{\bf e}_{5})$ (which depends on the choice of $\xi$) but the value of
$\lambda_{{\bf e}_{5}}$. A regular basis will thus be a standard five-vector
basis where all ${\bf e}_{\alpha} \in {\cal Z}$ and ${\bf e}_{5} = {\bf i}$.

\vspace{2ex} \begin{flushleft} \bf
2. Five-vectors as operators \\
\vspace*{2.5ex}
\rm A. \it Another representation for five-vectors
\end{flushleft}
In modern textbooks on differential geometry, ordinary tangent vectors are
usually introduced by identifying their fields with linear differential
operators (derivations) that act upon scalar functions from a certain set
$\Im$ which determines the topological and differential properties of the
manifold. Each derivation is a map
\begin{displaymath}
{\bf U}: \Im \rightarrow \Im
\end{displaymath}
that satisfies the following requirements:
\begin{equation} \left. \begin{array}{l}
{\bf U}[k] = 0 \mbox{ for any constant function } k \in \Im, \\ {\bf U}[f+g]
= {\bf U}[f] + {\bf U}[g] \mbox{ for any } f,g \in \Im, \\ {\bf U}[fg] =
{\bf U}[f] \cdot g + f \cdot {\bf U}[g]\mbox{ for any } f,g \in \Im.
\end{array} \right. \end{equation}
One can then prove a theorem that in a local coordinate system each
derivation can be presented as the following differential operator:
\begin{equation}
{\bf U} = U^{\alpha} (\partial/\partial x^{\alpha}),
\end{equation}
where $\partial/\partial x^{\alpha}$ are derivatives along coordinate lines
and $U^{\alpha}$ are scalar functions from $\Im$. It is evident that at
each point in space-time there exists a natural isomorphism between the
equivalence classes of parametrized curves corresponding to relation (2)
and operators of the form (10):
\begin{displaymath}
{\bf A} \mapsto \partial_{\bf A},
\end{displaymath}
and basing on this isomorphism one can identify the elements of $V_{4}$ with
four-dimensional tangent vectors.

Let us now find a similar operator representation for five-vectors. First,
one should notice that the two conditions that determine the equivalence
relation (3) can be replaced with a single requirement: that for any scalar
function $f$
\begin{displaymath}
\partial_{\cal A} f |_{Q} + \lambda_{\cal A}(Q) f(Q)
= \partial_{\cal B} f |_{Q} + \lambda_{\cal B}(Q) f(Q).
\end{displaymath}
This enables one to establish a one-to-one correspondence between the
equivalence classes of parametrized curves associated with relation (3)
and differential-algebraic operators of the form
\begin{equation}
{\bf u} = u^{\alpha} (\partial/\partial x^{\alpha}) + u^{5} \cdot {\bf 1},
\end{equation}
where $\bf 1$ is the identity operator. The simplest variant of such a
correspondence is evidently
\begin{equation}
{\bf a} \mapsto \partial_{\bf a} + \lambda_{\bf a} \cdot {\bf 1}.
\end{equation}

One can then consider five-vector fields and basing on the above
correspondence, relate them to such maps ${\bf u}: \Im \rightarrow \Im$
which in any local coordinate system can be presented in the form (11),
where $u^{A}$ are now scalar functions.

Finally, one can find a set of formal requirements, similar to conditions
(9) for derivations, that enable one to introduce the above maps without
referring to any coordinates. One possible set of such requirements is the
following:
\begin{equation} \left. \begin{array}{l}
{\bf u}[k] = \upsilon \cdot k \mbox{ for any constant } k \in \Im, \\
\; \mbox{ where } \upsilon \in \Im \mbox{ is characteristic of } {\bf u}, \\
{\bf u}[f+g] = {\bf u}[f] + {\bf u}[g] \mbox{ for any } f,g \in \Im, \\
{\bf u}[fg]={\bf u}[f] \cdot g + f \cdot {\bf u}[g] - {\bf u}[{\it 1}] fg
\mbox{ for} \\ \; \mbox{ any } f,g \in \Im, \mbox{ where } {\it 1}
\mbox{ is the constant} \\ \; \mbox{ unity function.}
\end{array} \right. \end{equation}
It is evident that any operator of the form (11) satisfies these three
requirements. Let us now prove the reverse statement:
\begin{quote}
In any local coordinate system each map ${\bf u}: \Im \rightarrow \Im$
satisfying requirements (13) can be presented in the form (11), where
$u^{A}$ are scalar functions from $\Im$.
\end{quote}
{\em Proof} : Let us consider the operator
\begin{displaymath} \bf
w \equiv u - \upsilon \cdot 1,
\end{displaymath}
where $\upsilon$ is the scalar function from $\Im$ defined by the first of
the requirements (13). It is a simple matter to check that $\bf w$ satisfies
conditions (9) for derivations and therefore can be presented in any local
coordinate system as
\begin{displaymath}
{\bf w} = w^{\alpha} (\partial/\partial x^{\alpha}),
\end{displaymath}
where $w^{\alpha} \in \Im$. Consequently, in any such system $\bf u$ can be
presented in the form (11) with $u^{\alpha} = w^{\alpha}$ and $u^{5} =
\upsilon. \; \rule{0.8ex}{1.7ex}$

One may observe that the operator corresponding to a given four-vector
$\bf U$ is exactly the differential part of the operator that corresponds to
any five-vector belonging to $\bf U$. This coincidence is a manifestation of
the fact that $V_{4}$ is isomorphic to $\cal Z$. This does not mean, however,
that one can identify four-vectors with $\cal Z$-components of five-vectors,
for as one will see in section 3, the isomorphism between $V_{4}$ and
$\cal Z$ is not preserved by parallel transport.

The representation of five-vectors with operators enables one to introduce
the former in another way: as maps $\Im \rightarrow \Im$ that satisfy
requirements (13). In its mathematical qualities, such a definition of
five-vectors is superior to the one given in section 1 and enables one to
introduce in a natural way the commutator of two five-vector fields. On the
other hand, in this case one cannot see as clearly the correspondence
between five-vectors and parametrized curves, and this is why in this
paper I have first considered the representation of five-vectors in the
form of equivalence classes associated with relation (3). It turns out,
however, that one should make a distinction between a given equivalence
class and the five-vector corresponding to it. In view of this, in the
following five-vector fields will always be identified with operators
satisfying requirements (13), the set of which will be denoted as $\FF$.

As in the case of four-vectors, tangent five-vectors at a given point $Q$
can be defined as equivalence classes of maps from $\FF$ with respect to
the equivalence relation
\begin{displaymath}
{\bf u} \equiv {\bf v} \Leftrightarrow {\bf u}[\, f \, ](Q) =
{\bf v}[\, f \, ](Q) \mbox{ for any } f \in \Im.
\end{displaymath}
The algebraic properties of the five-vectors defined this way are the same
as of those defined as classes of equivalent curves, and their analysis
would have been almost an exact repetition of the one made in section 1,
except for a few obvious changes in the definitions. Let me only mention
that a {\em regular} five-vector basis can now be defined as a basis where
all ${\bf e}_{\alpha}$ are purely differential operators and ${\bf e}_{5}
= {\bf 1}$.

One should also note that the correspondence between equivalence classes
of parametrized curves and operators from $\FF$ given by formula (12) is
not the only one possible. A more general form of such a correspondence is
\begin{equation}
{\bf a} \mapsto a \cdot \partial_{\bf a} + b \cdot \lambda_{\bf a} \cdot
{\bf 1},
\end{equation}
where $a$ and $b$ are some nonzero coefficients independent of $\bf a$.
Since the overall normalization of the operators representing five-vectors
is of no importance, one can always choose it so that $a=1$. In formula
(12) the second coefficient has been selected in the simplest way: $b=1$.
However, as one will see in section 3, to give a consistent definition to
the five-vectors associated with curves parametrized by {\em dimensional}
parameters, one has to assign to $b$ a certain dimension, so it will equal
unity only at some particular choice of the corresponding measurement units.

\vspace{2ex} \begin{flushleft}
B. \it Commutator of five-vector fields
\end{flushleft}
The representation of five-vectors with operators enables one to introduce
the commutator of five-vector fields. Namely, if ${\bf u} = u^{\alpha}
(\partial / \partial x^{\alpha}) + u^{5} \cdot {\bf 1}$ and ${\bf v} =
v^{\alpha}\ (\partial/\partial x^{\alpha}) + v^{5} \cdot {\bf 1}$, then by
definition,
\begin{displaymath}
[{\bf u,v}](f) = {\bf u}({\bf v}(f)) - {\bf v}({\bf u}(f))
\end{displaymath}
for any scalar function $f$, and one can show that $\bf w \equiv [u,v]$ is
an operator of the form (11) with components
\begin{equation}
w^{A} = u^{\beta} (\partial v^{A}/\partial x^{\beta}) - v^{\beta}
(\partial u^{A} / \partial x^{\beta}).
\end{equation}
For an arbitrary five-vector basis ${\bf e}_{A}$ one can define the
commutation constants, $C_{AB}^{\hspace*{1em} D}$, as
\begin{displaymath}
[{\bf e}_{A},{\bf e}_{B}] = C_{AB}^{\hspace*{1em} D} {\bf e}_{D},
\end{displaymath}
and show that the components of $[{\bf u,v}]$ in this basis are
\begin{displaymath}
\partial_{\bf u} v^{A}-\partial_{\bf v} u^{A}+u^{B} v^{D}
C_{BD}^{\hspace*{1em} A}.
\end{displaymath}
This is the analog of the well-known formula for components of the
commutator of two four-vector fields $\bf U$ and $\bf V$ in an arbitrary
basis ${\bf E}_{\alpha}$:
\begin{displaymath}
\partial_{\bf U} V^{\mu}-\partial_{\bf V} U^{\mu}+U^{\alpha} V^{\beta}
C_{\alpha \beta}^{\; \; \; \; \mu},
\end{displaymath}
where $[{\bf E}_{\alpha},{\bf E}_{\beta}] =
C_{\alpha \beta}^{\; \; \; \; \mu} {\bf E}_{\mu}$.

If ${\bf e}_{A}$ is a standard basis, one has
$C_{\mu 5}^{\hspace*{1em} \alpha} = 0$. It is a simple matter to
show that if $\bf u \in U$ and $\bf v \in V$, then $[{\bf u,v}] \in
[{\bf U,V}]$. Thus, if ${\bf e}_{\alpha} \in {\bf E}_{\alpha}$, then
$C_{\alpha \beta}^{\hspace*{1em} \mu} |_{\mbox{\scriptsize for
five-vectors}} = C_{\alpha \beta}^{\hspace*{1em} \mu} |_{\mbox{\scriptsize
for four-vectors}}$.

Let us now consider two subsets of five-vector fields from $\FF$: $(i)$
the subset $\FF_{\! \cal Z}$ of all purely differential operators, and
$(ii)$ the subset $\FF_{\! \cal E}$ of all purely algebraic operators. It is
evident that any element of $\FF$ can be uniquely presented as a sum of an
operator from $\FF_{\! \cal Z}$ and an operator from $\FF_{\! \cal E}$, so
$\FF = \FF_{\! \cal Z} \oplus \FF_{\! \cal E}$. The components of an
arbitrary five-vector field $\bf u$ in these two subspaces will be denoted as
${\bf u}^{\cal Z}$ and ${\bf u}^{\cal E}$. It is evident that they correspond
to the operators $u^{\alpha} (\partial/\partial x^{\alpha})$ and $u^{5} \cdot
{\bf 1}$, respectively.

One can easily see that the commutator of two five-vector fields from
$\FF_{\! \cal Z}$ is, again, a field from $\FF_{\! \cal Z}$, so
$\FF_{\! \cal Z}$ is a subalgebra: $[\FF_{\! \cal Z}, \FF_{\! \cal Z}]
\subset \FF_{\! \cal Z}$. Furthermore, the commutator of a field from
$\FF_{\! \cal E}$ with any other field from $\FF$ is an element of
$\FF_{\! \cal E}$, so $\FF_{\! \cal E}$ is an ideal:
$[\FF_{\! \cal E}, \FF] \subset \FF_{\! \cal E}$.

\vspace{3ex}

Commutators of four-vector fields enable one to tell whether or not a given
four-vector basis is holonomic. Namely, for a given set of basis fields
${\bf E}_{\alpha}$ there exists a system of local coordinates $x^{\alpha}$
such that ${\bf E}_{\alpha}$ are tangent vectors to coordinate lines
(${\bf E}_{\alpha} = \partial/\partial x^{\alpha}$) iff $[{\bf E}_{\alpha},
{\bf E}_{\beta}] = 0$. A similar statement for five-vectors is the following:
\begin{quote}
For a given set of standard five-vector basis fields ${\bf e}_{A}$
there exists a system of local coordinates $x^{\alpha}$ such that
${\bf e}_{\alpha}$ are tangent five-vectors to coordinate lines iff
\begin{flushright}
\hspace*{5ex} \hfill $[{\bf e}^{\cal Z}_{\alpha},{\bf e}^{\cal Z}_{\beta}]
= 0,$ \hspace*{4ex} \hfill {\rm (16a)} \\ \vspace*{1ex}
\hspace*{5ex} \hfill $[{\bf e}^{\cal Z}_{\alpha},{\bf e}^{\cal E}_{\beta}]
= \delta_{\alpha \beta} \cdot {\bf 1}.$ \hfill {\rm (16b)}
\end{flushright}
where $\delta_{\alpha \beta}$ is the Kronecker symbol.\footnote{For
simplicity, this theorem is formulated and proved for $a = b = 1$ in
formula (14).}
\end{quote} \setcounter{equation}{16}
{\em Proof} : If ${\bf e}_{\alpha}$ are tangent vectors to coordinate lines
$x^{\alpha}$, then ${\bf e}_{\alpha} = \partial / \partial x^{\alpha}
+ x^{\alpha} \cdot {\bf 1}$, and equations (16) are evidently obeyed.

If ${\bf e}_{\alpha}$ satisfy equations (16) and ${\bf E}_{\alpha}$ are
such that ${\bf e}_{\alpha} \in {\bf E}_{\alpha}$, then
\begin{displaymath} \left. \begin{array}{l}
0 = [{\bf e}^{\cal Z}_{\alpha},{\bf e}^{\cal Z}_{\beta}]
= \partial_{{\bf e}_{\alpha}} \partial_{{\bf e}_{\beta}}
- \partial_{{\bf e}_{\beta}} \partial_{{\bf e}_{\alpha}} \\
\hspace{10ex} = \partial_{{\bf E}_{\alpha}} \partial_{{\bf E}_{\beta}}
- \partial_{{\bf E}_{\beta}} \partial_{{\bf E}_{\alpha}}
= [{\bf E}_{\alpha},{\bf E}_{\beta}],
\end{array} \right. \end{displaymath}
and by virtue of the corresponding theorem for four-vectors, there exists
a system of local coordinates $x^{\alpha}$ such that $\partial/\partial
x^{\alpha} = \partial_{{\bf E}_{\alpha}} = \partial_{{\bf e}_{\alpha}}$.
In these coordinates each $\lambda_{{\bf e}_{\alpha}}$ is a certain real
function, which according to (16b) satisfies the equation
\begin{displaymath}
\partial \lambda_{{\bf e}_{\beta}}(x) / \partial x^{\alpha}
= \delta_{\alpha \beta}.
\end{displaymath}
This is only possible if $\lambda_{{\bf e}_{\alpha}}(x) = x^{\alpha} +
c^{\alpha}$, where $c^{\alpha}$ are integration constants. Consequently, one
has ${\bf e}_{\alpha} = \partial/\partial y^{\alpha} + y^{\alpha} \cdot {\bf
1}$, where $y^{\alpha} = x^{\alpha} + c^{\alpha}. \; \rule{0.8ex}{1.7ex}$

By analogy with four-vectors, a standard five-vector basis satisfying
requirements (16) can be called a {\em coordinate} basis. In certain cases,
however, it proves to be more convenient to select the $\cal E$-components
of the first four basis five-vectors in a different way, for example, equal
to zero. Since such five-vector bases still correspond to a coordinate
four-vector basis, it makes sense to call them coordinate, too.

\vspace{2ex} \begin{flushleft}
C. \it Five-vector Lie derivative
\end{flushleft}
The formal definition of the Lie derivative with respect to a four-vector
field $\bf U$ is the following:

$\bullet$ the Lie derivative of a four-vector field $\bf V$ is
\begin{equation}
\Lie_{\bf U} \bf V \equiv [U,V];
\end{equation}

$\bullet$ the Lie derivative of a scalar function $f$ is
\begin{equation}
\Lie_{\bf U} f \equiv {\bf U} f;
\end{equation}

$\bullet$ the Lie derivatives of all other four-tensor fields can be found
from formulae (17) and (18) by using the Leibniz rule, which in schematic
form can be presented as
\begin{equation}
\Lie_{\bf U} ({\cal A} \ast {\cal B}) = \Lie_{\bf U} {\cal A} \ast {\cal B}
+ {\cal A} \ast \Lie_{\bf U} {\cal B},
\end{equation}
where $\cal A$ and $\cal B$ are any two four-tensor fields and $\ast$
denotes contraction or tensor product.

In a similar manner one can give a formal definition to the Lie derivative
with respect to a {\em five}-vector field $\bf u$. I will denote this
latter derivative as $\Lie_{\bf u}$ and will call it the {\em five-vector
Lie derivative}. The analog of rule (17) is quite apparent:

$\bullet$ the five-vector Lie derivative of a five-vector field $\bf v$ is
\begin{equation}
\Lie_{\bf u} \bf v \equiv [u,v].
\end{equation}
As the analog of rule (18) it seems reasonable to take the following one:

$\bullet$ the five-vector Lie derivative of a scalar function $f$ is
\begin{equation}
\Lie_{\bf u} f \equiv {\bf u} f.
\end{equation}
It is easy to check that the five-vector Lie derivative of the product of
two scalar functions and the five-vector Lie derivative of the product of
a scalar function and a five-vector field are expressed in terms of the
five-vector Lie derivatives of the factors not according to the Leibniz
rule but according to the rule
\begin{equation}
\Lie_{\bf u} ({\cal A} \ast {\cal B}) = \Lie_{\bf u} {\cal A} \ast {\cal B}
+ {\cal A} \ast \Lie_{\bf u} {\cal B} - \Lie_{\bf u} {\it 1} \cdot
 ({\cal A} \ast {\cal B}),
\end{equation}
where, as before, $\it 1$ is the constant unity scalar function. In view
of this, it is not clear which of the rules --- (19), (22) or some other
--- should hold for the contraction and tensor product. To answer this
question and to gain a better understanding of the five-vector Lie
derivative, let us find for the latter an interpretation similar to
the one that can be given to the ordinary Lie derivative in terms of the
one-parameter local group of diffeomorphisms generated by a four-vector
field.

Let us recall that any sufficiently smooth four-vector field $\bf U$ defines
in the neighbourhood of any point $Q$ of the space-time manifold $\cal M$ a
congruence of integral curves, and that there always exist such an open
neighbourhood $\cal U$ of $Q$ and such a real number $\varepsilon > 0$ that
the map $\phi_{t}$ obtained by taking each point of $\cal U$ a parametric
distance $t$ along the corresponding integral curve, at $|t| < \varepsilon$
is a diffeomorphism of $\cal U$ into $\cal M$. At sufficiently small $s$ and
$t$ one has $\phi_{s} \circ \phi_{t} = \phi_{s+t}$ and $(\phi_{t})^{-1} =
\phi_{-t}$, so these diffeomorphisms form a one-parameter local group.

At each $t$ map $\phi_{t}$ defines a certain transformation, $\Phi_{t}$,
of scalar functions: the image $\Phi_{t} \{ f \}$ of a scalar function $f$
is such that
\begin{equation}
\Phi_{t} \{ f \} |_{\phi_{t}(P)} = f|_{P}.
\end{equation}
This transformation, in its turn, generates a certain transformation
of four-vector and other four-tensor fields, which is determined by the
following rules:

$\bullet$ the image $\Phi_{t} \{ {\bf V} \} $ of a four-vector field $\bf V$
is such that for any scalar function $f$
\begin{equation}
\Phi_{t} \{ {\bf V} \} \Phi_{t} \{ f \} = \Phi_{t} \{ {\bf V} f \};
\end{equation}

$\bullet$ the image $\Phi_{t} \{ \widetilde{\bf W} \} $ of a four-vector
1-form field $\widetilde{\bf W}$ is such that for any four-vector field
$\bf V$
\begin{equation}
< \Phi_{t} \{ \widetilde{\bf W} \}, \Phi_{t} \{ {\bf V} \} > \;
= \Phi_{t} \{ < \widetilde{\bf W}, {\bf V} > \} ;
\end{equation}

$\bullet$ the image $\Phi_{t} \{ {\cal A} \otimes {\cal B} \} $ of the
tensor product of two four-tensor fields $\cal A$ and $\cal B$ is such that
\begin{equation}
\Phi_{t} \{ {\cal A} \otimes {\cal B} \} = \Phi_{t} \{ {\cal A} \} \otimes
\Phi_{t} \{ {\cal B} \} . \vspace*{2ex}
\end{equation}
Within this approach, the Lie derivative of an arbitrary four-tensor field
$\cal S$ is defined as
\begin{equation}
\Lie_{\bf U} {\cal S} \equiv - \, (d/dt) \Phi_{t} \{ {\cal S} \} |_{t=0} .
\end{equation}
It is easy to see that at small $t$
\begin{equation}
\Phi_{t} \{ f \} = f - t \cdot {\bf U} f + O(t^2),
\end{equation}
from which, using definition (27), one obtains rule (18). In a similar
manner, after rewriting equation (24) as
\begin{displaymath}
\Phi_{t} \{ {\bf V} \} f = \Phi_{t} \{ {\bf V} \Phi_{t}^{-1} \{ f \} \}
\end{displaymath}
and using definition (27), one obtains rule (17). From equation (25) it
follows that the Leibniz rule holds for the contraction of a four-vector
field and a four-vector 1-form field and from equation (26) it follows that
it also holds for the tensor product of any two four-tensor fields. Thus,
the definition of the Lie derivative by means of equations (23)--(27) is
equivalent to its formal definition according to equations (17)--(19).

It is now apparent that to obtain the desired interpretation of the
five-vector Lie derivative, one should associate with every sufficiently
smooth five-vector field a certain one-parameter group of transformations of
scalar functions and five-tensor fields. Let us denote the transformations
from this group as $\Psi_{t}$ and define the five-vector Lie derivative of
an arbitrary five-tensor field $\cal S$ as
\begin{equation}
\Lie_{\bf u} {\cal S} \equiv - \, (d/dt) \Psi_{t} \{ {\cal S} \} |_{t=0} .
\end{equation}
Considering what has been said above, it seems reasonable to take that at
small $t$
\begin{equation}
\Psi_{t} \{ f \} = f - t \cdot {\bf u} f + O(t^2)
\end{equation}
for any scalar function $f$, which together with definition (29) gives us
rule (21). If, by analogy with rule (24), one then takes that
\begin{equation}
\Psi_{t} \{ {\bf v} \} \Psi_{t} \{ f \} = \Psi_{t} \{ {\bf v} f \}
\end{equation}
for any $\bf v$ and $f$, from formulae (29) and (30) one will obtain rule
(20). Thus, the infinitesimal transformation (30) produces the desired
result. Let us now find the corresponding finite transformation.

It is evident that for any sufficiently smooth five-vector field $\bf u$,
in the vicinity of any point $Q$ one can construct a congruence of integral
curves of the corresponding four-vector field $\bf U$. In this case these
curves will be called the integral curves of field $\bf u$. It is not
difficult to prove that at finite $t$ the image $\Psi_{t} \{ f \}$ of any
scalar function $f$ of class ${\rm C}^{\infty}$ equals
\begin{equation}
\Psi_{t} \{ f \} (\lambda) = \exp \{ - \int^{\lambda}_{\lambda-t}
u^{5}(\lambda') \, d\lambda' \, \} \, f(\lambda - t),
\end{equation}
where $\lambda$ is the parameter of the integral curve of field $\bf u$ and
$u^{5}$ is the fifth component of the latter in a regular basis. We thus see
that transformation $\Psi_{t}$ consists in ``shifting'' every value of the
function a parametric distance $t$ along the corresponding integral curve
and then multiplying it by a certain exponential factor. It is easy
to see that this latter factor equals the corresponding value of
$\Psi_{t} \{ {\it 1} \}$, so for an arbitrary scalar function $f$ one has
\begin{equation}
\Psi_{t} \{ f \} = \Psi_{t} \{ {\it 1} \} \Phi_{t} \{ f \}.
\end{equation}
>From the latter formula it follows that transformations $\Phi_{t}$ induced
by four-vector fields are a particular case of transformations $\Psi_{t}$
--- a case that corresponds to the five-vector fields from $\FF_{\! \cal Z}$.
Another particular case are the transformations $\Psi_{t}$ induced by
five-vector fields from $\FF_{\! \cal E}$. In this case
\begin{displaymath}
\Psi_{t} \{ f \} (P) = \exp \{ - \, t \cdot u^{5}(P) \} f(P).
\end{displaymath}

It is evident that to each transformation $\Psi_{t}$ one can put into
correspondence a certain map of $\cal U$ into $\cal M$, namely, the map
$\phi_{t}$ induced by the four-vector field corresponding to $\bf u$. Thus,
both in the case of four-vector fields and in the case of five-vector fields
one is actually dealing with {\em two} maps: $(i)$ a map from $\cal U$
to $\cal M$ and $(ii)$ a map from the set of restrictions to $\cal U$
of all the functions from $\Im$ to the set of restrictions of all these
functions to $\phi_{t}({\cal U})$. In the case of four-vector fields there
exists a one-to-one correspondence bewteen these two maps, which enables
one to think that the second map is induced by the first one. This is not
so in the case of five-vector fields: for example, the identity map from
$\cal U$ to $\cal M$ may correspond to different {\em non}identical
transformations of scalar functions.

>From equation (32) it is not difficult to derive that for any two scalar
functions $f$ and $g$
\begin{equation}
\Psi_{t} \{ fg \} = \Phi_{t} \{ f \} \Psi_{t} \{ g \}
= \Psi_{t} \{ f \} \Phi_{t} \{ g \},
\end{equation}
so in the general case the image of the product of two scalar functions with
respect to $\Psi_{t}$ is not the product of their images. By substituting
$\Psi^{-1}_{t} \{ {\it 1} \} \Psi_{t} \{ f \}$ for $\Phi_{t} \{ f \}$ in
formula (34) and differentiating both sides of the latter with respect to
$t$, one can verify that in this case rule (22) is indeed obeyed.

It is natural to define the action of $\Psi_{t}$ on a tensor product in the
following way:
\begin{equation}
\Psi_{t} \{ {\cal A} \otimes {\cal B} \} = \Psi_{t} \{ {\cal A} \} \otimes
\Psi_{t} \{ {\cal B} \} ,
\end{equation}
where $\cal A$ and $\cal B$ are any two five-tensor fields of nonzero rank.
This formula does not work, however, if one of the fields or both of them
are of rank zero. In the second case this can be seen from formula (34), if
one considers that for scalar functions $f \otimes g = fg$. In the first
case, if, for example, ${\cal A} = f$ and ${\cal B} = {\bf v}$, from formula
(34) and definition (31) one can easily obtain that
\begin{equation}
\Psi_{t} \{ f \otimes {\bf v} \} = \Psi_{t} \{ f {\bf v} \} =
\Phi_{t} \{ f \} \Psi_{t} \{ {\bf v} \} .
\end{equation}

Difficulties also occur with the definition of the action of $\Psi_{t}$ on
five-vector 1-forms. The direct analog of rule (25) is
\begin{equation}
< \Psi_{t} \{ \widetilde{\bf w} \}, \Psi_{t} \{ {\bf v} \} > \;
= \Psi_{t} \{ < \widetilde{\bf w}, {\bf v} > \},
\end{equation}
which means that the operation of contraction is ``correlated'' with
transformation $\Psi_{t}$ in the sense that the contraction of the image
of a five-vector field $\bf v$ with the image of a five-vector 1-form
field $\widetilde{\bf w}$ equals the image of the scalar function equal
to the contraction of $\bf v$ with $\widetilde{\bf w}$. The quantity
$< \widetilde{\bf w}, {\bf v} >$ can also be regarded as a five-tensor field
of rank zero obtained by contracting the field $\widetilde{\bf w} \otimes
{\bf v}$ of rank $(1,1)$. A similar operation can be performed on other
five-tensor fields, for example, on the field $\widetilde{\bf w} \otimes
{\bf v} \otimes {\bf s}$. For the contraction of this latter field to be
correlated with $\Psi_{t}$ it is necessary that there would hold not rule
(37) but the rule
\begin{equation}
< \Psi_{t} \{ \widetilde{\bf w} \}, \Psi_{t} \{ {\bf v} \} > \;
= \Phi_{t} \{ < \widetilde{\bf w}, {\bf v} > \}.
\end{equation}
Therefore, in those cases where $\Psi_{t}$ does not coincide with $\Phi_{t}$,
the requirements of correlation between the contraction and transformation
$\Psi_{t}$ for five-tensor fields of rank $(1,1)$ and for five-tensor fields
of other ranks are conflicting.

It is also useful to look at the components of the five-vector Lie
derivatives of five-tensor fields of different ranks, in a regular
coordinate basis. Let us write out these components for the case where
the rule that determines the action of $\Psi_{t}$ on 1-form fields is
\begin{equation}
< \Psi_{t} \{ \widetilde{\bf w} \}, \Psi_{t} \{ {\bf v} \} > = ( \Psi_{t}
\{ {\it 1} \} )^{k} \Psi_{t} \{ < \widetilde{\bf w}, {\bf v} > \}.
\end{equation}
According to equation (21), the five-vector Lie derivative of function $f$ is
\begin{equation}
\Lie_{\bf u} f = u^{\alpha} \partial_{\alpha} f + u^{5} f.
\end{equation}
>From equations (20) and (15) one finds that the components of the five-vector
Lie derivative of a five-vector field $\bf v$ are
\begin{equation}
(\Lie_{\bf u} {\bf v})^{A} = u^{B} (\partial_{B} v^{A})
- v^{B} (\partial_{B} u^{A}),
\end{equation}
where, for convenience, I have introduced the notation $\partial_{A} \equiv
\partial_{{\bf e}_{A}}$, so $\partial_{\alpha} = \partial / \partial
x^{\alpha}$ and $\partial_{5} = 0$. From equation (39) one can easily derive
that in the dual basis of five-vector 1-forms $\widetilde{\bf o}^{A}$,
\begin{equation} \left. \begin{array}{l}
(\Lie_{\bf u} \widetilde{\bf w})_{A} = u^{B} (\partial_{B} w_{A})
+ w_{B} (\partial_{A} u^{B}) \\ \hspace{25ex} \rule{0ex}{3ex}
+ (1+k) \, u^{5} w_{A}
\end{array} \right. \end{equation}
Finally, in the general case of an arbitrary five-tensor field
\begin{displaymath}
{\bf T} = T^{A_{1} \ldots A_{m}}_{B_{1} \ldots B_{n}} \;
{\bf e}_{A_{1}} \otimes \ldots \otimes {\bf e}_{A_{m}} \otimes
\widetilde{\bf o}^{B_{1}} \otimes \ldots \otimes \widetilde{\bf o}^{B_{n}}
\end{displaymath}
one has
\begin{equation} \left. \begin{array}{rcl}
(\Lie_{\bf u} {\bf T})^{A_{1} \ldots A_{m}}_{B_{1} \ldots B_{n}} & = &
u^{H} (\partial_{H} T^{A_{1} \ldots A_{m}}_{B_{1} \ldots B_{n}}) \\ & + &
\rule{0ex}{3ex} n (1+k) \cdot u^{5} \, T^{A_{1} \ldots A_{m}}_{B_{1} \ldots
B_{n}} \\ & - & \rule{0ex}{3ex} T^{H \ldots A_{m}}_{B_{1} \ldots B_{n}}
(\partial_{H} u^{A_{1}}) \\ - \; \ldots & - & \rule{0ex}{3ex} T^{A_{1}
\ldots H}_{B_{1} \ldots B_{n}} (\partial_{H} u^{A_{m}}) \\ & + &
\rule{0ex}{3ex} T^{A_{1} \ldots A_{m}}_{H \ldots B_{n}} (\partial_{B_{1}}
u^{H}) \\ + \; \ldots & + & \rule{0ex}{3ex} T^{A_{1} \ldots A_{m}}_{B_{1}
\ldots H} (\partial_{B_{n}} u^{H}).
\end{array} \right. \end{equation}

As one can see from the formulae obtained, there exists a distinguished value
of parameter $k$: $k=-1$, at which the terms proportional to $u^{5}$ in
equations (42) and (43) vanish, and the five-vector Lie derivative of any
five-tensor field that has at least one lower index depends only on the
{\em derivative} of $u^{5}$, as does the five-vector Lie derivative of a
five-vector field. One can also see that in the case of a five-tensor field
of rank zero (at $m=n=0$) formula (43) disagrees with formula (40) for the
five-vector Lie derivative of a scalar function.

All these observations suggest that in the case of transformations $\Psi_{t}$
induced by five-vector fields, one should make a distinction between scalar
functions which are elements of $\Im$ and scalar functions which are
five-tensor fields of rank zero. Formally, these two types of objects are of
different nature: the former are the functions upon which act the operators
of five-vector fields; the latter are elements of a commutative ring, by
which one can multiply five-vector fields, obtaining five-vector fields
again. To establish order in the theory, one should suppose that these two
types of functions are transformed by $\Psi_{t}$ {\em differently}: the
elements of $\Im$ are transformed according to formula (32), whereas the
five-tensor fields of rank zero are transformed according to the formula
\begin{equation}
\Psi_{t} \{ {\rm f} \} (\lambda) = {\rm f} (\lambda - t),
\end{equation}
which means that for them transformation $\Psi_{t}$ coincides with $\Phi_{t}$.
Under this assumption formula (35) for the tensor product will be valid for
five-tensor fields of zero rank as well. Moreover, since the contraction of
a vector and a 1-form is a tensor of rank zero, formula (37) will coincide
with formula (38), and consequently the contraction will be correlated with
transformation $\Psi_{t}$ for tensor fields of any rank for which it makes
sense. Among other things, the latter two facts mean that the five-vector
Lie derivative of a contraction and of a tensor product is expressed in
terms of the five-vector Lie derivatives of the factors according to the
Leibniz rule. In formulae (42) and (43) one should now put $k=-1$, and so
the derivatives $\Lie_{\bf u}$ of the corresponding five-tensor fields will
depend only on the derivative of $u^{5}$. Finally, the five-vector Lie
derivative of an arbitrary five-tensor field $\rm f$ of rank zero will be
\begin{equation}
\Lie_{\bf u} {\rm f} = \partial_{\bf u} {\rm f}
= u^{\alpha} \partial_{\alpha} {\rm f},
\end{equation}
which agrees with formula (43). Let me emphasize once more that in the case
of scalar fields from $\Im$, the image of the product of two such functions
with respect to $\Psi_{t}$ will not equal the product of their images,
which is inevitable and has no relation to the definition of $\Psi_{t}$
for five-tensor fields.

\vspace{2ex} \begin{flushleft} \bf
3. Some other properties of five-vectors \\
\vspace*{2.5ex}
\rm A. \it  Parallel transport of five-vectors
\end{flushleft}
As for any other type of vector-like objects considered in space-time, one
can speak of parallel transport of five-vectors from one space-time point to
another. One can then define the covariant derivative of five-vector fields;
introduce the connection coefficients corresponding to a given five-vector
basis; construct the corresponding curvature tensor; etc. In doing all
this one does not have to use in any way the fact that five-vectors are
associated with space-time by their definition.

One should expect that the origin of five-vectors manifests itself in that
the rules of their parallel transport are related in some way to similar
rules for four-vectors and to the Riemannian geometry of space-time. It is
obvious that this relation cannot be derived from the algebraic properties
of five-vectors, and to obtain it one has to make some new assumptions about
five-vectors, which ought to be regarded as part of their definition.

Let us first consider the relation between the rules of parallel transport
for four- and five-vectors. The simplest and the most natural form of this
relation is obtained by postulating that parallel transport preserves the
algebraic relation between four- and five-vectors discussed in subsection
1.C. A more precise formulation of this statement is the following:
\begin{equation}  \begin{minipage}{40ex} \it
If four-vector $\bf U$ is the equivalence class of five-vector
$\bf u$, then the transported $\bf U$ is the equivalence class of the
transported $\bf u$.
\end{minipage} \end{equation}
This assumption is quite natural considering that $\bf u \in U$ means that
$\bf u$ and $\bf U$ correspond to the same direction in the manifold. It has
two consequences, which can be conveniently expressed in terms of connection
coefficients (the latter are defined in section 4 of part I).

Let us consider the parallel transport of vectors from an arbitrary point
$Q$ to a nearby point $Q'$. If two five-vectors at $Q$ belong to the same
equivalence class, then according to our assumption, the transported
five-vectors should also be equivalent. Since parallel transport is a linear
operation, this means that vectors from ${\cal E}_{\mbox{\scriptsize at }Q}$
are transported into vectors from ${\cal E}_{\mbox{\scriptsize at }Q'}$.
Consequently, in {\em any} standard five-vector basis,
\begin{equation}
G^{\alpha}_{\; 5 \mu} = 0.
\end{equation}

Let ${\bf e}_{A}$ be an arbitrary standard five-vector basis and let
${\bf E}_{\alpha}$ be the associated basis of four-vectors. If
${\bf E}_{\alpha}(Q)$ are transported into vectors ${\bf E}_{\beta}(Q')
C^{\beta}_{\,\alpha}$, then according to our assumption,
${\bf e}_{\alpha}(Q)$ should be transported into vectors
${\bf e}_{\beta}(Q') C^{\beta}_{\, \alpha} + {\bf e}_{5}(Q')
C^{5}_{\, \alpha}$, where the coefficients $C^{\beta}_{\, \alpha}$
are the same in both cases. This means that in the selected bases,
\begin{equation}
G^{\alpha}_{\; \beta \mu} = \Gamma^{\alpha}_{\; \beta \mu}.
\end{equation}

It is evident that assumption (46) tells one nothing about
$G^{5}_{\, \alpha \mu}$ and $G^{5}_{\, 5 \mu}$. To get an idea of what
these coefficients can be like, let us now consider a particular case
where the connection for five-vector fields is such that there exists a
certain local symmetry which can be formulated as the following principle:
\begin{equation}  \begin{minipage}{40ex}
For any set of scalar, five-vector and five-tensor fields defined in the
vicinity of any point $Q$ in space-time, by means of a certain procedure one
can construct a set of fields in the vicinity of any other point $Q'$, such
that at $Q'$ these new fields (which will be called {\em equivalent}) satisfy
the same algebraic and first-order differential relations that the original
fields satisfy at $Q$.
\end{minipage} \end{equation}
The procedure by means of which the equivalent fields are constructed can be
formulated as follows:
\begin{enumerate}
   \item Introduce at $Q$ a system of local Lorentz coordinates $x^{\alpha}$.
   \\ Introduce the corresponding {\em regular coordinate} five-vector basis
   ${\bf e}_{A}$. \\ Introduce the corresponding bases for all other
   five-tensors. \vspace*{-2ex}
   \item Each scalar field $f$ in the vicinity of $Q$ will then
   determine and be determined by one real coordinate function $f(x)$. \\
   Each five-vector field $\bf u$ in the vicinity of $Q$ will determine
   and be determined by five real coordinate functions $u^{A}(x)$
   ($=$ components of $\bf u$ in the basis ${\bf e}_{A}$). \\
   Each five-tensor field $\bf T$ in the vicinity of $Q$ will determine
   and be determined by an appropriate number of real coordinate functions
   $T^{AB \ldots C}_{DE \ldots F}(x)$ ($=$ components of $\bf T$ in the
   relevant tensor basis corresponding to ${\bf e}_{A}$). \vspace*{-2ex}
   \item Introduce at $Q'$ a system of local Lorentz coordinates
   $x'^{\alpha}$ such that $x'^{\alpha}(Q') = x^{\alpha}(Q)$. \\
   Introduce the corresponding regular coordinate five-vector
   basis ${\bf e}'_{A}$. \\
   Introduce the corresponding bases for all other five-tensors.
   \vspace*{-2ex}
   \item Then the equivalent scalar, five-vector and five-tensor fields in
   the vicinity of $Q'$ will be determined in coordinates $x'^{\alpha}$ and
   in the corresponding bases by the {\em same functions} $f(\cdot)$,
   $u^{A}(\cdot)$, \ldots , $T^{AB \ldots C}_{DE \ldots F}(\cdot)$ that
   determine the original fields in the vicinity of $Q$ in coordinates
   $x^{\alpha}$ and in the corresponding bases.
\end{enumerate}

At $Q'= Q$ the two mentioned systems of local Lorentz coordinates,
$x^{\alpha}$ and $x'^{\alpha}$, are related as follows:
\begin{displaymath} \left. \begin{array}{l}
x'^{\alpha}(P) = x^{\alpha}(Q) + \Lambda^{\alpha}_{\; \beta}
[x^{\beta}(P) - x^{\beta}(Q)] \\ \hspace{10ex} \rule{0ex}{3ex}
+ \mbox{ terms of order } [x^{\alpha}(P) - x^{\alpha}(Q)]^{3},
\end{array} \right. \end{displaymath}
where $P$ is an arbitrary point in the vicinity of $Q$ and $\Lambda^{\alpha}
_{\; \beta}$ is a matrix from O(3,1). Reasoning as in section 4 of part I,
one can show that in the regular basis associated with either of these
coordinate systems one should have $G^{\alpha}_{\; \beta \mu} (Q) =
G^{5}_{\; 5 \mu} (Q) = 0$ and $G^{5}_{\; \alpha \mu} (Q) \propto
\eta_{\alpha \mu}$. Since in these coordinates $g_{\alpha \beta} (Q) \propto
\eta_{\alpha \beta}$, too, this means that the connection coefficients
$G^{5}_{\; \alpha \mu}(Q)$ are proportional to the components of the metric
tensor. Denoting the proportionality factor as $-\varsigma$ and using the
obvious transformation formulae for five-vector connection coefficients, one
can show that in {\em any} regular five-vector basis
\begin{equation}
G^{5}_{\; 5 \mu} = 0 \hspace*{6ex} \vspace*{-2ex}
\end{equation}
and
\begin{equation}
G^{5}_{\; \alpha \mu} = - \varsigma g_{\alpha \mu}.
\end{equation}

>From requirement (49) it also follows that five-vector connection
coefficients should have the same form at any two points in space-time in
similar five-vector bases. In the case of four-vector connection coefficients
a similar condition is satisfied automatically, and therefore is not
necessary. For five-vectors this is a nontrivial requirement, which means
that $\varsigma$ in equation (51) should be a constant.

It is evident that the value of $\varsigma$ is not fixed by the symmetry
principle. Since for dimensionless coordinates and curve parameters the
connection coefficients are dimensionless and $g_{\alpha \beta}$ are measured
in the units of interval squared, $\varsigma$ should have the dimension
$(interval)^{-2}$. There is no sense in talking about five-vectors if
$\varsigma = 0$, for it is impossible to distinguish a five-vector with such
rules of parallel transport from a pair consisting of a four-vector and a
scalar. Indeed, $V_{5}$ is isomorphic to the direct sum of $V_{4}$ and the
space of scalars (regarded as one-dimensional vectors), and it is apparent
that at $\varsigma = 0$ this isomorphism is preserved by parallel transport.
Considering this, I will always assume that $\varsigma \neq 0$.

\vspace{2ex} \begin{flushleft}
B. \it Five-vectors associated with dimensional \\
   \hspace{2ex} curve parameters
\end{flushleft}
So far we have been dealing with dimensionless curve parameters and
coordinates. In practice, the latter are usually selected in such a way so
that their values would be associated in some particular way with certain
lengths, time intervals or angles determined by the space-time metric. For
example, any system of dimensionless Lorentz coordinates in flat space-time
is such that the square of the interval between any two events $A$ and $B$,
measured in certain units $\ell$, equals
\begin{displaymath} \left. \begin{array}{l}
[x^{0}(A) - x^{0}(B)]^{2} - [x^{1}(A) - x^{1}(B)]^{2} \\ \hspace{7ex}
\rule{0ex}{3ex} - \, [x^{2}(A) - x^{2}(B)]^{2} - [x^{3}(A) - x^{3}(B)]^{2}.
\end{array} \right. \end{displaymath}
It is evident that if one changes the unit for measuring the interval as
\begin{equation}
\ell \rightarrow k \ell \; \; (k > 0),
\end{equation}
the dimensionless Lorentz coordinates will change in the inverse proportion.
This enables one to consider the latter as numerical values of certain
dimensional quantities, $\bar{x}^{\alpha}$, measured in the units of
interval, and it is these latter quantities one usually has in mind when
using the term ``Lorentz coordinates''.

The situation is similar in all other cases and as in the above example,
enables one to introduce the corresponding dimensional coordinates. For
simplicity, in the following I will suppose that all four coordinates are
measured in the units of interval. A convenient property of such dimensional
coordinates is that the corresponding metric coefficients, defined by the
equation
\begin{displaymath}
d s^{2} = \bar{g}_{\alpha \beta} \, d \bar{x}^{\alpha} d \bar{x}^{\beta},
\end{displaymath}
are all dimensionless quantities. It is easy to see that $\bar{g}_{\alpha
\beta}$ are the values of the dimensional metric coefficients $g_{\alpha
\beta}$ that correspond to the dimensionless coordinates $x^{\alpha}$ which
are the values of $\bar{x}^{\alpha}$ at the given $\ell$.

The same idea can be used to define dimensional curve parameters (for
simplicity, let us consider only those of them which are measured in
the units of interval). One can then introduce the notion of a tangent
four-vector corresponding to a curve parameterized by a given dimensional
parameter $\bar{\lambda}$. Such four-vectors behave as dimensional
quantities in the sense that at each $\ell$ they have a certain ``value'',
which, by definition, is the four-vector that corresponds to the
dimensionless parameter $\lambda$ which is the value of $\bar{\lambda}$ for
the given $\ell$. The algebraic operations and parallel transport for such
dimensional four-vectors are defined on the basis of the corresponding
operations for four-vectors associated with dimensionless parameters. For
example, a sum of two dimensional four-vectors $\bf U$ and $\bf V$ is a
dimensional four-vector whose value at any $\ell$ equals the sum of the
corresponding values of $\bf U$ and $\bf V$. It is evident that when one
changes $\ell$ according to formula (52), the value of each dimensional
four-vector changes in the same proportion, owing to which the inner
product of any two such four-vectors is a dimensionless quantity. This
and other properties of four-vectors associated with dimensional curve
parameters are well known, and I will not discuss them any further.

Let us now see how one can define a tangent five-vector corresponding to
a curve para\-metrized by some dimensional parameter $\bar{\lambda}$.
Following the same idea that has been used for tangent four-vectors, one
should consider such a five-vector as a quantity that has a certain ``value''
at every choice of $\ell$. This ``value'' is the tangent five-vector that
corresponds to the dimensionless parameter $\lambda$ which is the value
of $\bar{\lambda}$ for the given $\ell$. Let us now find the operator that
corresponds to this latter five-vector.

According to section 2, the general form of the operator representing the
five-vector tangent to a curve parametrized by a given dimensionless
parameter $\lambda$ is
\begin{equation}
a \cdot d/d \lambda + b \cdot \lambda \cdot {\bf 1},
\end{equation}
where $a$ and $b$ are some arbitrary nonzero constants. As it has been
said above, the overall normalization of the operators representing
five-vectors can always be chosen in such a way that $a$ be unity. When
dimensionless curve parameters are considered by themselves---not as values
of some dimensional parameters, one can take $b=1$, too, as it has been
done in formula (12). However, if operator (53) represents the value of a
five-vector associated with a dimensional parameter $\bar{\lambda}$, the
value of $b$ has to depend on the choice of $\ell$. Indeed, let us suppose
that one has a dimensional five-vector, $\bf u$, represented by a purely
differential operator and one parallel transports it from a given
space-time point $Q$ to some other point $Q'$. By definition,
${\bf u}^{\rm transported}$ is the five-vector at $Q'$ whose value at any
$\ell$ equals the value of $\bf u$ at $Q$ transported from $Q$ to $Q'$ along
the selected path. It is evident that if one changes $\ell$ according to
formula (52), the value of $\bf u$ will change in the same proportion,
and since parallel transport is a linear operation, so will the value
of ${\bf u}^{\rm transported}$. Consequently, the algebraic part of
the operator representing ${\bf u}^{\rm transported}$, which in the
general case will not be zero, should change in the same proportion as
$\ell$, which is only possible if $b$ changes as $b \rightarrow k^{2} b$.

We thus see that in the case of five-vectors associated with dimensional
curve parameters, the coefficient $b$ in formula (53) has to be the value
of some nonzero constant with dimension $(interval)^{-2}$. Apart from being
nonzero, this constant is absolutely arbitrary, and it is convenient to
choose it equal to the constant $\varsigma$ introduced in the previous
subsection. The operator representing a five-vector associated with a
dimensional parameter $\bar{\lambda}$ can then be presented in the
following form:
\begin{equation}
d/d \bar{\lambda} + \bar{\lambda} \cdot \varsigma \cdot {\bf 1}.
\end{equation}

In a similar manner one can introduce five-vectors corresponding to
parameters with dimension other than that of the interval. The algebraic and
differential properties of all such five-vectors will be practically the same
as those of the five-vectors associated with dimensionless parameters, and
only the dimension of certain relevant quantities will be different. For
example, in the particular case considered above, both the inner product
$h'$ induced by the metric and the nondegenerate inner product $h$ are
dimensionless. The relation between the two is still given by formula (6),
only now $\xi$ has the dimension $(interval)^{-2}$.

In the case of dimensional five-vectors, there exist {\em three} convenient
ways to normalize the fifth basis vector in a standard five-vector basis
and, accordingly, there are three ways to define a regular basis.

In those cases where the emphasis is made on parallel transport of
five-vectors, it is convenient to choose ${\bf e}_{5} = \varsigma \cdot
{\bf 1}$. Then, in the corresponding regular basis (in the one where
the other four basis five-vectors belong to $\cal Z$) one will have
$G^{5}_{\; \alpha \mu} = - \, g_{\alpha \mu}$, and the fifth component
of any five-vector $\bf u$ will equal $\lambda_{\bf u}$. In the following,
such a basis will be referred to as an {\em active} regular basis.

In those cases where the emphasis is made on the action of five-vectors on
scalar functions, it is convenient to take ${\bf e}_{5} = {\bf 1}$. In the
corresponding regular basis one will then have $G^{5}_{\; \alpha \mu} = - \,
\varsigma g_{\alpha \mu}$ and the fifth component of any five-vector $\bf u$
will equal $\varsigma \lambda_{\bf u}$. In the following, such a basis will
be referred to as a {\em passive} regular basis.

Finally, in those cases where the emphasis is made on the inner product
of five-vectors (at some particular choice of $\xi$), it is convenient to
normalize ${\bf e}_{5}$ by the requirement $h({\bf e}_{5},{\bf e}_{5}) =
{\rm sign} \, \xi$. It is evident that this equation has two solutions:
${\bf e}_{5} = + \, |\xi|^{-1/2} \, \varsigma \cdot {\bf 1}$ and
${\bf e}_{5} = - \, |\xi|^{-1/2} \, \varsigma \cdot {\bf 1}$, and to be
definite, I will choose the first one. In the corresponding regular basis
one will then have $G^{5}_{\; \alpha \mu} = - \, |\xi|^{1/2} g_{\alpha \mu}$,
and the fifth component of any five-vector $\bf u$ will equal $|\xi|^{1/2}
\lambda_{\bf u}$. In the following, such a basis will be referred to as a
{\em normalized} regular basis and the operator $|\xi|^{-1/2} \, \varsigma
\cdot {\bf 1}$ will be denoted as $\bf n$.

>From now on, unless it is stated otherwise, I will talk only about
five-vectors associated with dimensional curve parameters and coordinates,
and will omit the bar over the dimensional $x^{\alpha}$ and $\lambda$. It
is evident that any result obtained for such five-vectors can readily be
reformulated for five-vectors corresponding to dimensionless parameters.

\vspace{2ex} \begin{flushleft}
C. \it Four-vectors as simple bivectors over $V_{5}$
\end{flushleft}
We are now ready to demonstrate that the five-vectors introduced formally in
part I can be identified with the five-dimensional tangent vectors introduced
in this paper. More precisely, it will be shown that there can be established
a natural isomorphism between the space of four-vectors and one of the maximal
vector spaces of simple bivectors over $V_{5}$ and that in those cases where
the connection for five-vectors possesses the local symmetry considered in
subsection A, this isomorphism is preserved by parallel transport. This will
mean that the five-vectors considered in this paper have all the formal
properties postulated for five-vectors in part I.

Let us fix a nonzero five-vector ${\bf e} \in {\cal E}$ and consider all
simple bivectors of the form ${\bf u} \wedge {\bf e}$, where ${\bf u} \in
V_{5}$. It is evident that ${\bf u} \wedge {\bf e} = {\bf v} \wedge {\bf e}$
if and only if ${\bf u} - {\bf v} \in {\cal E}$, which is exactly the
equivalence relation $R$ of subsection 1.C. Thus, one is able to establish
a one-to-one correspondence between four-vectors and elements of the maximal
vector space of simple bivectors over $V_{5}$ with the directional vector
belonging to $\cal E$. It is evident that this correspondence is a
homomorphism and that it depends on the choice of the arbitrary nonzero
vector $\bf e$. Let us fix the latter by requiring that the considered
correspondence be an isomorphism.

Let us consider some particular nondegenerate inner product on $V_{5}$,
where the constant $\xi$ has been chosen positive, so that $h$ would have
the signature $(+---+)$. It is not difficult to check that if $\bf u \in U$
and $\bf v \in V$, then
\begin{equation}
g({\bf U,V}) = h({\bf u,v}) - \frac{h({\bf e,u}) h({\bf e,v})}{h({\bf e,e})}.
\end{equation}
On the other hand, the inner product of ${\bf u} \wedge {\bf e}$ and
${\bf v} \wedge {\bf e}$ induced by $h$ is
\begin{displaymath}
h({\bf u} \wedge {\bf e},{\bf v} \wedge {\bf e}) = h({\bf u,v}) h({\bf e,e})
- h({\bf u,e}) h({\bf v,e}).
\end{displaymath}
For the correspondence ${\bf U} \mapsto {\bf u} \wedge {\bf e}$ to be an
isomorphism $g({\bf U,V})$ should equal $h({\bf u} \wedge {\bf e}, {\bf v}
\wedge {\bf e})$ for all $\bf u$ and $\bf v$, which is only possible if
$h({\bf e,e}) = 1$. This means that $\bf e$ is either $\bf +n$ or
$\bf -n$. We thus see that (for the given $\xi > 0$) there exist {\em two}
isomorphisms of $V_{4}$ onto the considered maximal vectors space of simple
bivectors, and unless additional requirements are imposed, the choice between
the two is a matter of convention. To be definite, I will take $\bf e = n$.

The fact that the above isomorphism (actually, both of them) is preserved by
parallel transport becomes evident if one considers that the relation $\bf u
\in U$ is invariant under parallel transport and that $\bf n$ is transported
into $\bf n$.

One can now use all the results obtained within the formal theory of
five-vectors. Most of the definitions made in the present paper correspond
to those made in part I. The only essential difference concerns the
associated four-vector basis.

When introducing five-vectors formally, one has no means of associating them
with four-dimensional tangent vectors other than saying that a five-vector
$\bf u$ corresponds to the four-vector identified with the bivector ${\bf u}
\wedge {\bf e}$, where $\bf e$ is some directional vector. The only way one
can fix $\bf e$ within the formal theory is to require that it be of certain
length. However, since the inner product of five-vectors is an object of
study itself, one prefers to have a purely ``kinematic'' relation between
the four- and five-vector bases, and the only sensible choice is to take
${\bf E}_{\alpha} = {\bf e}_{\alpha} \wedge {\bf e}_{5}$. This means that
\begin{equation}
{\bf E}_{\alpha} = \xi^{1/2} \lambda_{{\bf e}_{5}} \times (\mbox{the
equivalence class of } {\bf e}_{\alpha}).
\end{equation}
Considering that $\xi \, (\lambda_{{\bf e}_{5}})^{2} = h({\bf e}_{5},
{\bf e}_{5})$, from formula (55) one obtains the relation between the
components of $g$ and $h$ derived in part I:
\begin{displaymath}
g_{\alpha \beta} = h_{55} h_{\alpha \beta} - h_{\alpha 5} h_{\beta 5}.
\end{displaymath}
Furthermore, if $\nabla_{\mu}{\bf n} = 0$, then $G^{5}_{\; 5 \mu} =
(\lambda_{{\bf e}_{5}})^{-1} \partial_{\mu} \lambda_{{\bf e}_{5}}$, and for
the four-vector connection coefficients corresponding to basis (56) one has
\begin{displaymath}
\Gamma^{\alpha}_{\, \beta \mu} = G^{\alpha}_{\, \beta \mu}
+ \delta^{\alpha}_{\beta} G^{5}_{\, 5 \mu},
\end{displaymath}
which is exactly the relation obtained in part I. Finally, if one
assumes that flat space-time possesses the symmetry considered in
subsection A, then in any orthonormal standard five-vector basis one will
have
\begin{displaymath}
G^{5}_{\, 5 \mu} = 0 \; \mbox{ and } \; G^{5}_{\, \alpha \mu} = - \, \kappa
\eta_{\alpha \mu},
\end{displaymath}
where $\kappa = \xi^{1/2}$ (if we had taken $\bf e = - n$, we would have
had $\kappa = - \, \xi^{1/2}$).

Let me also say a few words about the equation for the first covariant
derivative of $h$. Straightforward calculations similar to those made in
part I give the following result:
\begin{equation} \begin{array}{l}
\{ \nabla_{\bf U} h \} ({\bf v,w}) \\ \hspace{8ex} = \kappa g({\bf U,V})
h({\bf w,n}) \\ \hspace{16ex} + \; \kappa g({\bf U,W}) h({\bf v,n}),
\end{array} \end{equation}
where it is assumed that $\bf v \in V$ and $\bf w \in W$. Since for any $\bf
v$ one has $\kappa h({\bf v,n}) = \xi \cdot \lambda_{\bf v}$, this equation
can also be presented as
\begin{equation}
\{ \nabla_{\bf U} h \} ({\bf v,w}) = \xi \, g({\bf U,V}) \lambda_{\bf w}
 + \xi \, g({\bf U,W}) \lambda_{\bf v}.
\end{equation}
It is easy to see that for an arbitrary nonzero ${\bf e} \in {\cal E}$
the bivectors $\bf v \wedge e$ and $\bf w \wedge e$ correspond to
the four-vectors $\xi^{1/2} \lambda_{\bf e} {\bf V}$ and $\xi^{1/2}
\lambda_{\bf e} {\bf W}$, respectively. Thus, by multiplying both sides of
equation (57) by $\xi \, (\lambda_{\bf e})^{2} = h({\bf e,e})$ one obtains
\begin{displaymath} \begin{array}{l}
h({\bf e,e}) \{ \nabla_{\bf U} h \} ({\bf v,w}) \\
\hspace{8ex} = \kappa g({\bf U}, {\bf v} \wedge {\bf e}) h({\bf w,e}) \\
\hspace{16ex} + \; \kappa g({\bf U}, {\bf w} \wedge {\bf e}) h({\bf v,e}),
\end{array} \end{displaymath}
which is exactly the equation for $\nabla h$ obtained within the formal
theory of five-vectors.

\vspace{2ex} \begin{flushleft}
D. \it Operator $\nabla$ and matrix $g$ with five-vector indices
\end{flushleft}
Above I have introduced the covariant derivative operator, $\nabla_{\bf U}$,
which differentiates five-vector fields in the direction specified by its
argument---by the four-vector $\bf U$. As a consequence, the corresponding
connection coefficients, $G^{A}_{\, B \mu}$, have indices of two kinds: two
five-vector indices $A$ and $B$ and one four-vector index $\mu$. This is not
very convenient in those cases where indices of different kinds have to be
permuted, for any relation with such permutations is valid only if the four-
and five-vector bases have been chosen accordingly.

This inconvenience can be easily eliminated if instead of $\nabla_{\bf U}$
one considers the operator $\nabla_{\bf u}$, defined by the relation
\begin{equation}
\nabla_{\bf u} = \nabla_{\bf U} \; \mbox{ for } \; \bf u \in U.
\end{equation}
It is obvious that $\nabla_{\bf u}$ is absolutely equivalent to $\nabla_{\bf
U}$. However, unlike the latter, it formally depends on a {\em five}-vector.
It is evident that $\nabla_{\bf u} = \nabla_{({\bf u}^{\cal Z})}$ for
any five-vector $\bf u$, so for any ${\bf e} \in {\cal E}$ one has
$\nabla_{\bf e} = 0$. Operator $\nabla_{\bf u}$ is the analog of the
operator $\partial_{\bf u}$ that acts upon scalar functions, and
relation (59) is the analog of the relation
\begin{displaymath}
\partial_{\bf u} = \partial_{\bf U} \; \mbox{ for } \; \bf u \in U.
\end{displaymath}

It is natural to introduce the notation $\nabla_{A} \equiv
\nabla_{{\bf e}_{A}}$. Then, in any standard five-vector basis one has
$\nabla_{5} = 0$ and $\nabla_{\mu}^{\mbox{\scriptsize (with a five-vector
index)}} = \nabla_{\mu}^{\mbox{\scriptsize (with a four-vector index)}}$.
In view of this, I will use the same carrier letter `$G$' to denote the
connection coefficients corresponding to $\nabla_{A}$:
\begin{displaymath}
\nabla_{A} {\bf e}_{B} = {\bf e}_{C} \, G^{C}_{\; BA}.
\end{displaymath}
Then $G^{A}_{\: B \mu}$ with a five-vector $\mu$ will equal
$G^{A}_{\: B \mu}$ with a four-vector $\mu$ in any standard basis, and
rules (47), (48), (50), and (51) will apply to $G^{A}_{\; BC}$ without
any changes. In addition, one will have a fifth rule: that in any standard
five-vector basis,
\begin{displaymath}
G^{A}_{\; B5} = 0.
\end{displaymath}

In the usual manner one can derive the transformation formula for
$G^{A}_{\; BC}$, corresponding to the basis transformation
${\bf e}'_{A} = {\bf e}_{B} L^{B}_{\, A}$:
\begin{displaymath}
G'^{A}_{\; BC} = (L^{-1})^{A}_{\; D} G^{D}_{\; EF} L^{E}_{\; B} L^{F}_{\; C}
+ (L^{-1})^{A}_{\; D} (\partial_{F} L^{D}_{\; B}) L^{F}_{\; C}.
\end{displaymath}
If both bases are standard, one will have $G'^{A}_{\; B 5} = G^{A}_{\; B 5}
= 0$ and
\begin{displaymath}
G'^{A}_{\; B \mu} = (L^{-1})^{A}_{\; D} G^{D}_{\; E \nu} L^{E}_{\; B}
L^{\nu}_{\; \mu} + (L^{-1})^{A}_{\; D} (\partial_{\nu} L^{D}_{\; B})
L^{\nu}_{\; \mu},
\end{displaymath}
which is the usual formula for transformation of connection coefficients.

In a similar manner one can deal with four-vector indices in $g_{\mu \nu}$.
Actually, I have already defined the corresponding five-vector quantity in
subsection 1.F, where it has been denoted as $h'$. From now on, instead of
$h'({\bf u,v})$ I will use the notation $g({\bf u,v})$, so formulae (5)
and (6) will acquire the form:
\begin{displaymath}
g({\bf u,v}) = g({\bf U,V})
\end{displaymath}
for $\bf u \in U$ and $\bf v \in V$, and
\begin{equation}
h({\bf u,v}) = g({\bf u,v}) + \xi \cdot \lambda_{\bf u} \lambda_{\bf v}.
\end{equation}

It is evident that $g({\bf u,v}) = g({\bf u}^{\cal Z},{\bf v}^{\cal Z})$ for
any five-vectors $\bf u$ and $\bf v$, so for any ${\bf e} \in {\cal E}$ one
has $g({\bf u,e}) = 0$. If one now introduces the notation $g_{AB} \equiv
g({\bf e}_{A},{\bf e}_{B})$, then in any standard five-vector basis one will
have
\begin{displaymath}
g_{55} = g_{\alpha 5} = g_{5 \alpha} = 0 \vspace*{-2ex}
\end{displaymath}
and \vspace*{1ex}
\begin{displaymath}
g_{\alpha \beta}^{\mbox {\scriptsize (with five-vector indices)}} =
g_{\alpha \beta}^{\mbox {\scriptsize (with four-vector indices)}}.
\end{displaymath}
>From these formulae and equations (47) and (48) of subsection A it follows
that in any standard five-vector basis
\begin{displaymath}
\partial_{\mu} g_{AB} - g_{CB}G^{C}_{\; A \mu} - g_{AC} G^{C}_{\; B \mu} = 0,
\end{displaymath}
which means that $g$ {\em regarded as a five-tensor} satisfies the equation
$\nabla g = 0$.

The latter equation and formula (60) enable one to obtain the following
expression for the first covariant derivative of the inner product $h$
regarded as a five-tensor:
\begin{displaymath}
\{ \nabla_{\bf u} h \} ({\bf v,w}) = \xi \, \{ \nabla_{\bf u}
\lambda \}_{\bf v} \lambda_{\bf w} + \xi \, \lambda_{\bf w} \{
\nabla_{\bf u} \lambda \}_{\bf v},
\end{displaymath}
where $\{ \nabla_{\bf u} \lambda \}_{\bf v} \equiv \partial_{\bf u}
\lambda_{\bf v} - \lambda_{(\nabla_{\bf u} {\bf v})}$. Comparing this
expression with equation (58), one can see that the latter is equivalent
to the following simpler equation:
\begin{equation}
\{ \nabla_{\bf u} \lambda \}_{\bf v} = g({\bf u,v}).
\end{equation}

\vspace{2ex} \begin{flushleft}
E. \it Forms associated with five-vectors
\end{flushleft}
As in the case of any other type of vectors, one can consider linear forms
corresponding to five-vectors. Such forms will be denoted with lower-case
boldface Roman letters with a tilde: $\widetilde{\bf a}$, $\widetilde{\bf b}$,
$\widetilde{\bf c}$, etc., and their space will be denoted as
$\widetilde{V}_{5}$. To distinguish a $p$-form associated with five-vectors
from a $p$-form associated with four-vectors I will call the former a
{\em five-vector} $p$-form and the latter a {\em four-vector} $p$-form.

Five-vector 1-forms have all the properties common to linear forms in
general. In addition, they have several specific features, which are due
to their association with five-vectors, and it is these latter properties
I will now consider.

The existence of two distinguished subspaces in $V_{5}$ results in the
existence of two distinguished subspaces in $\widetilde{V}_{5}$. The first
of these subspaces is made up by all those 1-forms from $\widetilde{V}_{5}$
whose contraction with any five-vector from ${\cal E}$ is zero. It is
evident that this subspace is four-dimensional, and I will denote it as
$\widetilde{\cal Z}$. The other distinguished subspace is made up by all
those 1-forms that have a zero contraction with any five-vector from
${\cal Z}$. This subspace is one-dimensional, and I will denote it as
$\widetilde{\cal E}$. It is easy to see that $\widetilde{\cal Z}$ and
$\widetilde{\cal E}$ have only one common element---the zero 1-form, and
that $\widetilde{V}_{5}$ is the direct sum of $\widetilde{\cal Z}$ and
$\widetilde{\cal E}$. The components of an arbitrary five-vector 1-form
$\widetilde{\bf w}$ in these two subspaces will be denoted as
$\widetilde{\bf w}^{\widetilde{\cal Z}}$ and
$\widetilde{\bf w}^{\widetilde{\cal E}}$, respectively.

If ${\bf e}_{A}$ is a standard five-vector basis and $\widetilde{\bf o}^{A}$
is the corresponding dual basis of five-vector 1-forms, then
$\widetilde{\bf o}^{\alpha} \in \widetilde{\cal Z}$ for all $\alpha$.
The fifth basis 1-form will not necessarily be an element of
$\widetilde{\cal E}$: this will be the case only if all ${\bf e}_{\alpha}
\in {\cal Z}$. The same conclusions follow from the transformation formulae
for the dual basis of 1-forms, corresponding to the transformation
${\bf e}'_{A} = {\bf e}_{B} L^{B}_{\; A}$ from one standard five-vector
basis to another. Since in this case $(L^{-1})^{\alpha}_{\; 5} = 0$, one has
\begin{displaymath}
\widetilde{\bf o}'^{\; \alpha} = (L^{-1})^{\alpha}_{\; B} \;
\widetilde{\bf o}^{B} = (L^{-1})^{\alpha}_{\; \beta} \;
\widetilde{\bf o}^{\beta},
\vspace*{-2ex} \end{displaymath}
but
\begin{displaymath}
\widetilde{\bf o}'^{\; 5} = (L^{-1})^{5}_{\; 5} \; \widetilde{\bf o}^{5} +
(L^{-1})^{5}_{\; \beta} \; \widetilde{\bf o}^{\beta}. \hspace*{2ex}
\end{displaymath}
If ${\bf e}_{A}$ is a passive regular basis, then $\widetilde{\bf o}^{5} \in
\widetilde{\cal E}$ and $<\widetilde{\bf o}^{5},{\bf 1}> \; = 1$. This
particular five-vector 1-form will be denoted as $\jj$.

The fact that $\cal Z$ is isomorphic to $V_{4}$ enables one to establish a
natural isomorphism between $\widetilde{\cal Z}$ and the space of four-vector
1-forms, which will be denoted as $\widetilde{V}_{4}$. Namely, to each
five-vector 1-form $\widetilde{\bf w}$ from $\widetilde{\cal Z}$ one can put
into correspondence such a four-vector 1-form $\widetilde{\bf W}$ that for
any five-vector ${\bf u} \in {\cal Z}$ one will have $\bf <\widetilde{w},
u> \; = \; <\widetilde{W}, U>$, where $\bf u \in U$. It is evident that
this isomorphism can be extended to a map of $\widetilde{V}_{5}$ onto
$\widetilde{V}_{4}$, which will be a homomorphism but will not be a
one-to-one correspondence. In the standard way, this latter map defines
an equivalence relation on $\widetilde{V}_{5}$:
\begin{equation}
\widetilde{\bf u} \equiv \widetilde{\bf v} \; \mbox{ iff their images in }
\widetilde{V}_{4} \mbox{ are equal}.
\end{equation}
This enables one to regard $\widetilde{V}_{4}$ as a quotient set and
four-vector 1-forms as equivalence classes. It is not difficult to see that
the equality of the images of $\widetilde{\bf u}$ and $\widetilde{\bf v}$ in
$\widetilde{V}_{4}$ is equivalent to $\widetilde{\bf u} - \widetilde{\bf v}
\in \widetilde{\cal E}$. The relation between $\widetilde{V}_{4}$ and
$\widetilde{V}_{5}$ is thus similar to the relation between $V_{4}$ and
$V_{5}$, however, unlike the latter, it is not preserved by parallel
transport, as it will be shown below.

The parallel transport of five-vector 1-forms is defined in the standard
way: by requiring that it conserve the contraction. Consequently, if
$G^{A}_{\; B \mu}$ are connection coefficients for a standard five-vector
basis, then for the corresponding dual basis of 1-forms one has
\begin{equation}
\nabla_{\mu}\widetilde{\bf o}^{A}=-\, G^{A}_{\; B \mu}\widetilde{\bf o}^{B},
\end{equation}
and from formulae (47) and (48) one obtains that
\begin{displaymath}
\nabla_{\mu} \widetilde{\bf o}^{\alpha}
= - \, G^{\alpha}_{\; B \mu} \widetilde{\bf o}^{B}
= - \, G^{\alpha}_{\; \beta \mu} \widetilde{\bf o}^{\beta}
= - \, \Gamma^{\alpha}_{\; \beta \mu} \widetilde{\bf o}^{\beta}.
\end{displaymath}
This means that 1-forms from $\widetilde{\cal Z}$ are transported into
1-forms from $\widetilde{\cal Z}$ and that the isomorphism between
$\widetilde{\cal Z}$ and $\widetilde{V}_{4}$ is preserved by parallel
transport. From formula (63) it also follows that
\begin{displaymath}
\nabla_{\mu} \widetilde{\bf o}^{5} = - \, G^{5}_{\; 5 \mu} \widetilde{\bf o}
^{5} - G^{5}_{\; \beta \mu} \widetilde{\bf o}^{\beta},
\end{displaymath}
which shows that in the general case, 1-forms from $\widetilde{\cal E}$ are
not transported into 1-forms from $\widetilde{\cal E}$, so equivalence
relation (62) is not invariant under parallel transport.

As in the case of any other vector space, each inner product on $V_{5}$
defines a certain correspondence between five-vectors and five-vector
1-forms. Since one has two inner products on $V_{5}$ --- $g$ and $h$,
there are two such correspondences, which I will denote as $\vartheta_{g}$
and $\vartheta_{h}$, respectively. By definition, $\vartheta_{g}(\bf u)$ is
such a five-vector 1-form that
\begin{equation}
< \! \vartheta_{g}({\bf u}),{\bf v} \! > \; = g \, ({\bf u,v}) \; \,
\mbox{ for any } \; {\bf v} \in V_{5}.
\end{equation}
The definition of the 1-form $\vartheta_{h}(\bf u)$ is similar. It is
evident that both $\vartheta_{g}$ and $\vartheta_{h}$ are linear maps
of $V_{5}$ into $\widetilde{V}_{5}$. If $u^{A}$ are components of some
five-vector $\bf u$ in a certain five-vector basis, then the components of
$\vartheta_{g}(\bf u)$ and $\vartheta_{h}(\bf u)$ in the corresponding dual
basis of 1-forms are $g_{AB} u^{B}$ and $h_{AB} u^{B}$, respectively. Since
the matrix $h_{AB}$ is nondegenerate, this means that $\vartheta_{h}$
is a one-to-one correspondence and is a map of $V_{5}$ {\em onto}
$\widetilde{V}_{5}$. It is also easy to see that $\vartheta_{h}(\cal Z) =
\widetilde{\cal Z}$ and $\vartheta_{h}(\cal E) = \widetilde{\cal E}$. By
contrast, $\vartheta_{g}$ is neither a one-to-one correspondence nor a
surjection. It is evident that $\vartheta_{g}({\bf u}) = \vartheta_{g}
({\bf u}^{\cal Z}) = \vartheta_{h}({\bf u}^{\cal Z})$, so $\vartheta_{g}
(\cal Z) = \widetilde{\cal Z}$, but $\vartheta_{g}(\cal E) = \{
\widetilde{\bf 0} \}$. Consequently, one can use $g_{AB}$ only to lower
five-vector indices. Raising indices with $g_{AB}$ is possible only if
one confines oneself to five-vectors from $\cal Z$ and to 1-forms from
$\widetilde{\cal Z}$.

All this is in agreement with the general theorem that asserts that the
following three statements are equivalent: $(i)$ the correspondence between
vectors and linear forms induced by a given inner product is injective;
$(ii)$ this correspondence is surjective; $(iii)$ the inner product is
nondegenerate.

Another general theorem states that the correspondence between vectors and
linear forms is invariant under parallel transport if and only if the
corresponding inner product is covariantly constant. Since $g$, as a
five-tensor, satisfies the equation $\nabla g = 0$, one has
\begin{displaymath}
[\vartheta_{g}({\bf u})]^{\rm transported}
= \vartheta_{g}({\bf u}^{\rm transported})
\end{displaymath}
for any $\bf u$. Alternatively, this can be expressed as
\begin{displaymath}
\nabla_{\bf v}[\vartheta_{g}({\bf u})] = \vartheta_{g}(\nabla_{\bf v}{\bf u})
\end{displaymath}
for all $\bf u$ and $\bf v$, which means that the lowering of five-vector
indices with $g_{AB}$ commutes with covariant differentiation.

As it has been discussed earlier, the nondegenerate inner product $h$ is
{\em not} covariantly constant, and so in the general case, $[\vartheta_{h}
({\bf u})]^{\rm transported}$ does not coincide with $\vartheta_{h}({\bf
u}^{\rm transported})$. Consequently, the lowering and raising of five-vector
indices with $h_{AB}$ does not commute with covariant differentiation, and
one should take special care whenever these two operations are performed on
the same five-tensor.

In section 5 of part I I have introduced the five-vector 1-form
$\widetilde{\bf x}$, which by definition coincides with the fifth element of
the 1-form basis dual to an active regular five-vector basis. Comparing this
with the definition of the 1-form $\jj$, one finds that $\widetilde{\bf x} =
\varsigma^{-1} \cdot \jj$. Furthermore, it is easy to see that for any
five-vector $\bf v$,
\begin{displaymath}
\lambda_{\bf v} = \; < \widetilde{\bf x} ,{\bf v} >.
\end{displaymath}
Substituting this expression for $\lambda_{\bf v}$ into the definition of
$\nabla \lambda$, one finds that
\begin{displaymath} \begin{array}{rcl}
\{ \nabla_{\bf u} \lambda \}_{\bf v} & = & \partial_{\bf u}
< \widetilde{\bf x} , {\bf v} > - < \widetilde{\bf x} , \nabla_{\bf u}
{\bf v} > \\ & = & < \nabla_{\bf u} \widetilde{\bf x} , {\bf v} >.
\end{array} \end{displaymath}
Substituting this latter expression and definition (64) into equation (61),
one obtains that
\begin{displaymath}
< \nabla_{\bf u} \widetilde{\bf x} - \vartheta_{g} ({\bf u}) \, , {\bf v} >
\; = 0
\end{displaymath}
for any five-vector $\bf v$, which means that
\begin{equation}
\nabla_{\bf u} \widetilde{\bf x} = \vartheta_{g}({\bf u})
\end{equation}
for any $\bf u$, which is nothing but equation (38) of part I. In
equation (65) the 1-form $\vartheta_{g}({\bf u})$ can be presented as a
contraction of $g$ regarded as a five-tensor of rank $(0,2)$, with the
five-vector $\bf u$. Considering also that $\nabla_{\bf u} \widetilde{\bf x}
= \; < \nabla \widetilde{\bf x} , {\bf u} >$, one can present equation (65)
as
\begin{displaymath}
\nabla \widetilde{\bf x} = g.
\end{displaymath}

Let me finally say a few words about five-vector $p$-forms with $p$
other than 1. It is a simple matter to see that any five-vector $p$-form
$\widetilde{\bf s}$ with $p > 1$ can be uniquely presented as a sum of two
terms: $(i)$ a $p$-form made only of 1-forms from $\widetilde{\cal Z}$
and $(ii)$ a wedge product of the type $\widetilde{\bf t} \wedge \jj$,
where $\widetilde{\bf t}$ is a $(p-1)$-form. In the following, these
two terms will be referred to as the $\widetilde{\cal Z}$- and
$\widetilde{\cal E}$-components of $\widetilde{\bf s}$, respectively,
and will be denoted as $\widetilde{\bf s}^{\widetilde{\cal Z}}$ and
$\widetilde{\bf s}^{\widetilde{\cal E}}$. It is easy to see that at $p = 1$
this definition agrees with the definition of the $\widetilde{\cal Z}$-
and $\widetilde{\cal E}$-components of a 1-form given above. It is obvious
that a five-vector 5-form has only the $\widetilde{\cal E}$-component, and
it is convenient to take that for any 0-form $f$,
\begin{displaymath}
f^{\widetilde{\cal Z}} = f \mbox{ and } f^{\widetilde{\cal E}} = 0.
\end{displaymath}
The application of five-vector forms in exterior differential calculus will
be discussed in detail in part IV.

\vspace{2ex} \begin{flushleft}
\bf Acknowledgement
\end{flushleft}
I would like to thank V. D. Laptev for supporting this work. I am grateful
to V. A. Kuzmin for his interest and to V. A. Rubakov for a very helpful
discussion and advice. I am indebted to A. M. Semikhatov of the Lebedev
Physical Institute for a very stimulating and pleasant discussion and to
S. F. Prokushkin of the same institute for consulting me on the Yang-Mills
theories of the de Sitter group. I would also like to thank L. A. Alania,
S. V. Aleshin, and A. A. Irmatov of the Mechanics and Mathematics Department
of the Moscow State University for their help and advice.

\end{document}